\documentclass[aps,amsmath,prd,floatfix
]{revtex4}

\usepackage{graphicx}

\newcommand{\gras}[1]{\mbox{\boldmath $#1$}}
\newcommand{\be}{\begin{equation}}
\newcommand{\ee}{\end{equation}}
\newcommand{\ba}{\begin{eqnarray}}
\newcommand{\ea}{\end{eqnarray}}
\newcommand{\Mc}{{\cal M}}
\newcommand{\Ms}{M_{\odot}}
\newcommand{\m}{\langle}
\newcommand{\M}{\rangle}

\newcommand{\bml}{\begin{mathletters}}
\newcommand{\eml}{\end{mathletters}}

\def\ltsima{$\; \buildrel < \over \sim \;$}
\def\simlt{\lower.5ex\hbox{\ltsima}}
\def\gtsima{$\; \buildrel > \over \sim \;$}
\def\simgt{\lower.5ex\hbox{\gtsima}}
\begin{document} 

\title{LISA observations of rapidly spinning
massive black hole binary systems}

\author{Alberto Vecchio}
%\email[]{av@star.sr.bham.ac.uk}
\affiliation{School of Physics and Astronomy, 
The University of Birmingham, Edgbaston, Birmingham B15 2TT, UK}

\date{\today}
\begin{abstract}

Binary systems of massive black holes will be detectable by the Laser Interferometer 
Space Antenna (LISA) throughout the entire Universe. Observations of
gravitational waves from this class of sources will have important repercussions on our
understanding of the behaviour of gravity in the highly non-linear relativistic regime,
the distribution and interaction of massive black holes at high redshift
and the formation and evolution of cosmic structures. It is therefore important
to address how accurately LISA can measure the source parameters and
explore the implications for astronomy and cosmology. 

Present observations and theoretical models suggest that massive black holes
could be spinning, possibly rapidly in some cases. 
In binary systems, the relativistic spin-orbit interaction 
causes the orbital plane to precess in space producing a characteristic signature
on the emitted gravitational waves.
In this paper we investigate the effect of spins on the gravitational wave signal registered
at the LISA output -- we provide ready-to-use analytical expressions of the
measured signal -- 
and the implications for parameter estimation.
We consider the in-spiral phase of binary systems in circular orbit 
undergoing the so-called "simple precession" and we approximate the
gravitational radiation at the restricted post$^{1.5}$-Newtonian order. 
We show that the presence of spins changes dramatically the signature of the signal
recorded by LISA. As a consequence, the mean square
errors associated to the parameter measurements are significantly smaller than the ones obtained
when the effect of spins is neglected. For a binary system of two $10^6\,\Ms$ black holes,
the angular resolution and the relative error on the luminosity distance improve by a 
factor $\approx$ 3-to-10; the fractional errors on the chirp mass and the reduced mass  
decrease by a factor $\sim 10$ and $\sim 10^3$, respectively.

\end{abstract}

\pacs{PACS numbers: 04.80.Nn, 95.55.Ym, 95.75.pq, 97.60.Lf}
\maketitle

\section{Introduction}
\label{sec:int}

Binary systems of massive black holes in the mass range $10^7\,\Ms - 10^5\,\Ms$ are 
among the most spectacular sources that will be observed by the Laser Interferometer 
Space Antenna (LISA), an ESA/NASA space-borne laser interferometer aimed at 
observations of gravitational waves (GW's) in the low-frequency region 
of the spectrum $\sim 10^{-5}$ Hz - 0.1 Hz, and expected to be launched in 
2011~\cite{lisa_ppa}.
The interest in the detection of gravitational radiation from massive black 
holes (MBH's) is two-fold~\cite{CT02}:
(i) for fundamental physics, because it will provide 
a thorough understanding of the behaviour of 
strong gravity in the highly non-linear relativistic regime, and (ii) for astronomy 
and cosmology because GW's provide an "orthogonal" and complementary mean, with respect
to conventional astronomy, of investigating the high-redshift formation and evolution of cosmic 
structures, such as galaxy interactions and mergers, and exploring 
the demography of massive black holes.

Massive black holes (MBH's) are considered to be essentially ubiquitous in the centre of 
galaxies in the local and low-redshift Universe~\cite{Richstoneetal98,Magorrianetal98}.
Our Galaxy~\cite{Genzeletal97} and NGC 
4258~\cite{Miyoshietal95} provide probably the most solid observational 
evidence for the existence of such black 
holes~\cite{Maoz98}. The evidence for the presence of black holes in galaxies
at high redshift is more circumstantial, but the fact that 
galaxies, at some stage of their life, harbour AGN's, and we observe
quasars at $z \simgt 6$~\cite{Fanetal01} strongly favour
the hypothesis that black holes are fairly common also at high redshift. This conclusion 
is supported by recent
results~\cite{WL02} that show that the high-redshift quasar distribution and the present 
day density
of black holes can be reconciled by invoking galaxy mergers where the relation 
between host velocity dispersion and black hole mass is consistent with the one observed 
locally~\cite{MF01}. The present day population of MBH's is therefore likely to be 
the dormant remnant of those earlier activities. Two black holes are brought together at the centre of
a merger remnant by dynamical
friction from the stars of the common host. The further
evolution of a binary must then take place on a time scale short enough for the system to
coalesce within an Hubble time due to radiation reaction. Two are the
main mechanisms that could make the binary hard enough to satisfy this condition: 
dynamical friction from stars in the core~\cite{BBR80,VCP94,RR95,MM01,Yu02} 
and gas dynamical effects~\cite{GR00,AN02}. However,
despite all these circumstantial evidences, the rate of coalescence of massive
black hole binaries in the Universe is still controversial: 
present astrophysical estimations vary widely from 
$\sim 0.1\,{\rm yr}^{-1}$ to $\sim 100\,{\rm yr}^{-1}$~\cite{BBR80,Haehnelt94,
Vecchio97,Haehnelt98,MHN01,Menou02,JB02,WL03}, depending on model assumptions and BH mass range. At present there are no direct observational evidences for the presence of
massive black holes binary systems, with the possible exception of OJ287~\cite{Pursimoetal00}.

LISA will be able to detect MBH binaries at a moderate-to-high 
signal-to-noise ratio throughout the entire Universe (redshift $z \approx 10$ and beyond, 
if black holes are already present). 
During the whole LISA observational campaign -- the nominal duration of the
mission is 3 years, but the instrument is expected to operate for
about 10 years, if no catastrophic failure occurs on the spacecraft --
one can therefore expect from a handful to a very large number of
detections.

LISA is an all-sky monitor. At any one time, the instrument maps the whole sky,
although with a different sensitivity depending on the location of the
source and the polarisation of the waves. By coherently tracking
the phase and amplitude evolution of GW's, one can extract precise information 
about a source. In observations of binary systems, the number of parameters 
on which the waveform depends, and that one needs to extract from the data, 
is large, in the most general case 17.
Clearly correlations among parameters are inevitable, degrading the precision 
of the measurements. 

A few studies have been carried out so far to address
how accurately LISA can measure the source parameters and the implications for 
astronomy and cosmology. Because of the complexity of the problem and open theoretical
issues in the modelling of the gravitational waveform in the most general case, 
assumptions, possibly ad-hoc, have been introduced on the waveform model.
Cutler~\cite{Cutler98} has considered the in-spiral phase of binary systems approximating the
waveform at the restricted post$^{1.5}$-Newtonian order. Sintes and collaborators~\cite{SV00,Sintes03}
and Moore and Hellings~\cite{MH02} have 
investigated the parameter estimation including post$^{2}$-Newtonian 
corrections to the amplitude and the phase. Hughes~\cite{Hughes02} has considered 
restricted  post$^{2}$-Newtonian waveforms for the in-spiral phase and has included
the ring-down portion of the signal; more recently Seto has included the effect
of the finite length of the LISA arms using 
restricted post$^{1.5}$-Newtonian waveforms~\cite{Seto02}.
All these analysis have {\em a priori} assumed that black
holes have no spins, or the spins are parallel to each other and 
the orbital angular momentum. 
However, present astrophysical observations and theoretical models suggest that 
massive black holes are likely to be rotating, possibly rapidly
in some cases. Observations of AGN's and quasars 
are highly consistent with rapidly rotating black holes;
observations of black holes in quiescence are in agreement
with a slow, if any, rotation and some theoretical models suggest 
small-to-moderate spins (see, {\em e.g.},~\cite{HB03}). 
If there is little doubt that black holes are, in general, rotating, how
strong spins are is to a large extent unknown.
Quite likely, gravitational wave observations shall provide the most 
effective tool in addressing this issue.
Spins are likely to add vast complexity to the signal recorded
at the detector output,
and require more parameters to describe the waveform~\cite{ACST94,Kidder95,VC98,Vecchio00}. 
In this paper we show 
that spins do change dramatically the structure of the signal observed by LISA 
and, as a consequence, affect significantly (even by orders of magnitudes for some
selected parameters) the errors with which LISA can measure the source parameters. 

In this paper we restrict our attention to the in-spiral phase of massive binary systems 
of comparable mass. We consider circular orbits, model
the waveform at the restricted post$^{1.5}$-Newtonian order and assume that binary systems
undergo the so-called {\it simple precession}~\cite{ACST94}, which is the relevant
scenario from an astrophysical viewpoint. The goal of the paper is two fold:
(i) to provide ready to use analytical
expressions for the LISA detector output when binary systems undergo spin-orbit
precession, and (ii) to show the dramatic change in parameter
estimation that occurs when spin-orbit modulation in phase and amplitude is
taken into account. 
The actual parameter space is so vast -- the waveform is described by 12 parameters --
that here we concentrate on typical systems
of two $10^6\,\Ms$ black holes. A thorough exploration of the whole parameter space 
requires large-scale, CPU-intensive Monte-Carlo simulations and is beyond the
scope of this paper; such analysis is currently 
in progress and will be reported in a separate paper~\cite{Vecchioetal03}.

The organisation of the paper is as follows. In section~\ref{sec:lisa} we 
review the main features of the LISA mission, in particular its orbital 
configuration which is central for the topic discussed in this paper. In 
Section~\ref{sec:gw} we review the main properties and key equations 
regarding the emission of GW's from in-spiralling binary systems, with emphasis
on the role plaid by spins, and we stress the differences with respect to 
the non-spinning, i.e. non-precessing, case. Section~\ref{sec:output} and~\ref{sec:error}
contain the key results of the paper. In Section~\ref{sec:output} we derive
explicit ready-to-use analytical expressions for the signal measured
at the LISA detector output for binary systems in circular orbit that 
undergo simple precession; the waveform is computed within the 
restricted post$^{1.5}$-Newtonian approximation. In section~\ref{sec:error} 
we explore the errors with which the
twelve parameters of the signal can be determined and compare the 
results with the case where spins are neglected. 
In Section~\ref{sec:concl} we summarise
our conclusions and present pointers to future work.

\section{The LISA mission}
\label{sec:lisa}

In this section we review the main properties of the LISA mission. We refer the
reader to~\cite{lisa_ppa} for more details.

\subsection{The LISA orbit}

LISA is an all-sky monitor with a quadrupolar antenna pattern. Its orbital
configuration is conceived in order to keep the geometry of the interferometer
as stable as possible during the mission, which in turn provides a thorough
coverage of the whole sky: a constellation 
of three drag-free spacecraft (containing the "free-falling test masses") 
is placed at the vertexes of an ideal equilateral triangle 
with sides $\simeq 5\times 10^6\,{\rm km}$; it forms
a three-arms interferometer, with a $60^\circ$ angle between two 
adjacent laser beams. The LISA orbital motion is as follows: the barycentre of 
the instrument follows an essentially circular heliocentric orbit,
$20^\circ$ behind the Earth; the detector plane is tilted by 
$60^\circ$ with respect to
the Ecliptic and the instrument counter-rotates around the normal to the 
detector plane with the same period $1\,{\rm yr}$. 

Following~\cite{Cutler98} we introduce two Cartesian reference frames~\cite{frame}:
(i) a fixed "barycentric" frame $(x,y,z)$ tied to the Ecliptic and centred
in the Solar System Barycentre, with ${\bf \hat z}$ perpendicular to 
the Ecliptic, and the plane $(x,y)$ in the Ecliptic itself; (ii) a
detector reference frame $(x',y',z')$, centred in the LISA centre of mass and 
attached to the detector, with ${\bf \hat z}'$ perpendicular to the plane
defined by the three arms and the $x'$ and $y'$ axis defined so that 
the unit vectors ${\bf \hat l}_j$ ($j = 1,2,3$) along each arm read
\be
{\bf \hat l}_j = \cos\left[\frac{\pi}{12} + \frac{\pi}{3}\,(j-1)\right] 
{\bf \hat x'}
+ \sin\left[\frac{\pi}{12} + \frac{\pi}{3}\,(j-1)\right] 
{\bf \hat y'}\,.
\label{larm}
\ee
In the Ecliptic frame the motion of LISA's centre-of-mass is described by the polar angles
\ba
\Theta  & = & \frac{\pi}{2}\,,\nonumber\\
\Phi(t) & = & \Phi_0 + n_{\oplus} t\,,
\label{barlisa}
\ea
where
\be
n_{\oplus} \equiv \frac{2\pi}{1\,{\rm yr}}\,,
\ee
and $\Phi_0$ sets the position of the detector at same arbitrary reference time. 
The normal to the detector plane ${\bf \hat z'}$ precesses around
${\bf \hat z}$ according to
\be
{\bf \hat z'} = \frac{1}{2} {\bf \hat z} -\frac{\sqrt{3}}{2}
\left[\cos\Phi(t) {\bf \hat x} + \sin\Phi(t) {\bf \hat y}\right]\,.
\label{zprimo}
\ee

The time evolution of the unit vectors ${\bf \hat l}_j$ ($j = 1,2,3$) 
along each arm is described by the following expression~\cite{Cutler98}:
\ba
{\bf \hat l}_j = & &
\left[\frac{1}{2}\,\sin\alpha_j(t)\,\cos\Phi(t) - \cos\alpha_j(t)\,\sin\Phi(t)\right]
\,{\bf \hat x} \nonumber\\
& + &
\left[\frac{1}{2}\,\sin\alpha_j(t)\,\sin\Phi(t) + \cos\alpha_j(t)\,\cos\Phi(t)\right]
\,{\bf \hat y} 
+ \left[\frac{\sqrt{3}}{2}\,\sin\alpha_j(t)\right]\,{\bf \hat z} \,,
\label{larm1}
\ea
where $\alpha_j(t)$ increases linearly with time, according to
\be
\alpha_j(t) = n_{\oplus} t - (j-1)\pi/3 + \alpha_0 \;,  
\label{alphaj}
\ee
and $\alpha_0$ is just a constant specifying the orientation of
the arms at the arbitrary reference time $t=0$.

\subsection{The LISA detector output}

The strain $h(t)$ produced at the output of the LISA Michelson interferometer
by a GW signal characterised by the two independent polarisation states 
$h_+(t)$ and $h_{\times}(t)$~\cite{Thorne87} is 
\be
h^{(\iota)}(t) = \frac{\sqrt{3}}{2}\,\left[
F^{(\iota)}_+(t) h_+(t) + F^{(\iota)}_{\times}(t) h_{\times}(t)
\right]\,.
\label{resp}
\ee
In Eq.~(\ref{resp}) $F^{(\iota)}_+(t)$ and $F^{(\iota)}_{\times}(t)$ are
the {\it time dependent} antenna patters and the factor $\sqrt{3}/2 = \sin(\pi/3)$ 
comes from the $60^\circ$ opening angle of the LISA arms. $F_+$ and $F_{\times}$
vary with time because during the observation the interferometer changes
orientation with respect to the source; in fact, in-spiral binaries
are long-lived sources in the low frequency band.
The index $\iota = I,II$ labels the two independent Michelson outputs that can be 
constructed from the readouts of the three arms if the noise is uncorrelated and 
"totally symmetric"~\cite{Cutler98}. They are equivalent to the outputs of
two identical interferometers in the same location, rotated by $45^\circ$ one
with respect to the other. 

The functions describing the interferometer beam pattern depend on the source
location in the sky ${\bf \hat N}$ (GW's propagate in the $-{\bf \hat N}$ 
direction) and wave polarisation, which is related to
the orientation of the orbital plane, and therefore the orbital angular momentum 
${\bf \hat L}$, with respect to the detector. With respect to the frame tied to 
the Solar system barycentre, ${\bf \hat N}$ and ${\bf \hat L}$
are described by the polar angles $(\theta_N\,,\phi_N)$ and $(\theta_L\,,\phi_L)$.
Equivalently, in the frame attached to the detector,  ${\bf \hat N}$ and ${\bf \hat L}$
are identified by the angles $(\theta_N'\,,\phi_N')$ and $(\theta_L'\,,\phi_L')$.
In the reference frame attached to LISA, the antenna beam patters read
\begin{subequations}
\be
F_+\left(\theta_N'\,,\phi_N'\,\psi_N'\right) =
\frac{1}{2} (1 + \cos\theta_N'^2) \cos 2\phi_N' \cos 2 \psi_N' -
\cos\theta_N' \sin 2\phi_N' \sin 2 \psi_N'\,, \\
\label{Fplus}
\ee
\be
F_{\times}\left(\theta_N'\,,\phi_N'\,\psi_N'\right) = 
\frac{1}{2} (1 + \cos\theta_N'^2) \cos 2\phi_N'\sin 2\psi_N'
+\cos\theta_N' \sin 2\phi_N' \cos 2 \psi_N'\,,
\label{Fcross}
\ee
\end{subequations}
where $\psi_N'$ describes the "polarisation angle" of the waveform in
the detector frame. The antenna patterns~(\ref{Fplus}) and~(\ref{Fcross}) for the
two outputs $\iota = I$ and $\iota = II$ are therefore
\ba
F_{+,\times}^{(\iota)}(t) & = &  
\left\{
\begin{array}{ll}
F_{+,\times}\left(\theta_N'\,,\phi_N'\,\psi_N'\right)\quad\quad & \quad\quad(\iota = I) \\
 & \\
F_{+,\times}\left(\theta_N'\,,\left(\phi_N'-\pi/4\right)\,,\psi_N'\right)
\quad\quad & \quad\quad (\iota = II) \\
\end{array}
\right.\,.
\label{FIII}
\ea
Note that because of the detector change of orientation,
the angles $\theta_N'$, $\phi_N'$ and $\psi_N'$ are time dependent.
As a function of the angles measured in the Solar System Barycentre they read
\ba
\cos\theta_N'(t) & = & \frac{1}{2} \cos\theta_N - \frac{\sqrt{3}}{2}
\sin\theta_N \cos(\Phi(t) - \phi_N) \,,\\
\label{thetaNl'}
\phi_N'(t) & = & \Xi_1 + \frac{\pi}{12} +
\tan^{-1}
\left\{\frac{\frac{\sqrt{3}}{2} \cos\theta_N + \frac{1}{2}
\sin\theta_N \sin(\Phi(t) - \phi_N)}
{\sin\theta_N \sin(\Phi(t) - \phi_N)}\right\}\,,\\
\label{phiNl'}
{\rm tan}\,\psi_N' & = & \frac{{\bf \hat L} \cdot {\bf \hat z}' - 
({\bf \hat L} \cdot {\bf \hat N})\,({\bf \hat z}' \cdot {\bf \hat N})}
{{\bf \hat N} \cdot ({\bf \hat L} \times {\bf \hat z}')} \,,
\label{psiNl'}
\ea
where
\be
\Xi_j = n_{\oplus} t - \frac{\pi}{12} - \frac{\pi}{3} (j-1) +\Xi_0\,,
\label{Xijl'}
\ee
and $\Xi_0$ sets the orientation of ${\bf \hat l}_j$ at $t = 0$. We are not yet 
ready to derive the explicit expression for $\psi_N'$, Eq.(\ref{psiNl'}),
because for spinning black holes, the source at the centre of this paper, ${\bf \hat L}$
is not a constant of motion. We derive the necessary equations in the following section.

\section{Gravitational waves from binary systems}
\label{sec:gw}

In this section we briefly review, partly to fix notation,
the basic concepts and formulae regarding the emission of gravitational waves
by binary systems, with emphasis on the effects produced by spins. 
We refer the reader to~\cite{ACST94,Kidder95} for details and to~\cite{Blanchet02} and
references therein for a thorough review regarding post-Newtonian gravitational waveforms.

The whole {\it coalescence} of a binary system is usually divided 
in three distinct phases~\cite{FH98}: (i) the adiabatic in-spiral, during which the 
BH orbital evolution occurs on the time-scale of the gravitational radiation reaction,
which is much longer than the orbital period; (ii) the merger, which takes place at
the end of the in-spiral when the binary orbit becomes relativistically unstable and
the black holes enter a free fall plunge;
and (iii) the ring-down, when the resulting BH settles down in its final stationary
Kerr state, oscillating according to its normal modes. 
In this paper we consider only the in-spiral phase. In fact the time to coalescence
for a system radiating at frequency $f$ is:
\be
\tau \simeq 1.2\times 10^7\, \left(\frac{f}{10^{-4}\,{\rm Hz}}\right)^{-8/3}\,
\left(\frac{M\,(1 + z)}{10^6\,\Ms}\right)^{-5/3}\,
\left(\frac{\eta}{0.25}\right)^{-1}\, {\rm sec}\,,
\ee
where $M$ and $\eta$ are the total mass and the symmetric mass ratio, respectively, defined
in Eqs.~(\ref{M})-(\ref{eta}), $\tau$ is the {\it observed} life-time of a binary system as
recorded in the solar system for a source at redshift $z$ with intrinsic total mass $M$ that 
enters the detector window at the observed frequency $f = 10^{-4}$ Hz. 
Binary systems are therefore long-lived in the LISA band, and
one critically relies on long integration times to disentangle the sources parameters, in 
particular to resolve the source position in the sky and measure its 
luminosity distance~\cite{Cutler98}. 

As we consider only the in-spiral phase of the coalescence, in the frequency domain we
shut-off the signal at frequency
\ba
f_{\rm isco} & = & \frac{1}{\pi\,6^{3/2} M\, (1 + z)} \nonumber\\
             & \simeq & 4.4 \times10^{-3}\, 
\left(\frac{M\,(1 + z)}{10^6\,\Ms}\right)^{-1}\,
{\rm Hz}\,,
\label{fisco}
\ea
which corresponds to the innermost stable circular orbit of a particle orbiting a 
Schwarzichild black hole (ISCO). The real transition from in-spiral to merger will
occur at a frequency somewhat (by a factor $\simlt 2$) different from $f_{\rm isco}$, 
but due to the difficulty of defining the ISCO we will adopt the value~(\ref{fisco}). The 
results presented in this paper are anyway essentially unaffected by changing the 
cut-off frequency by a small factor.

We also assume the orbit to be circular. This is completely reasonable from
an astrophysical point of view, as a massive binary system of roughly equal mass
black holes has enough time -- regardless of the initial eccentricity at which it
forms -- to circularise before it enters the observational window, due to dynamical
friction and radiation reaction~\cite{Peters64,VCP94}.

\subsection{The amplitude and phase evolution}
\label{subsec:Aphi}

We consider two massive black holes of mass $m_1$ and $m_2$, and spins ${\bf S}_1$ and 
${\bf S}_2$, respectively. Several mass parameters are actually useful:
\begin{subequations}
\ba
M   & = & m_1 + m_2 \,,\\
\label{M}
\mu & = & \frac{m_1 \, m_2}{M} \,, \\
\label{mu}
\Mc & = & \mu^{3/5}\,M^{2/5} = M\,\eta^{3/5} \,, \\
\label{Mc}
\eta & = & = \frac{\mu}{M} \,;\\
\label{eta}
\ea
\end{subequations}
they represent, respectively,  the total mass, the reduced mass, the chirp mass and
the symmetric mass ratio.

The system evolves by loosing energy and angular
momentum through emission of gravitational waves of increasing frequency
and amplitude. The emitted radiation is given by the superposition
of harmonics at multiples of the orbital period, and the two independent
polarisation amplitudes $h_{+}$ and $h_{\times}$~\cite{Thorne87} can 
be schematically represented as~\cite{BIWW96}
\be
h_{+,\times}(t) = \Re \left\{\sum_k H_{+,\times}^{(k)}(t)\,e^{i\,k\,\phi_{\rm orb}(t)}\right\}\,,
\label{hpc}
\ee
where $H_{+,\times}^{(k)}$ are time dependent functions of the source parameters and
$\phi_{\rm orb}(t)$ is the binary orbital phase. The strongest harmonic is associated
to the quadrupole moment of the source, and therefore corresponds to $k=2$. 
In this paper we consider the so-called {\it restricted post-Newtonian approximation} 
to the radiation~(\ref{hpc}): we retain only the leading order
contribution
to the amplitude, therefore only $H_{+,\times}^{(2)}$ at the Newtonian order, and
take into account post-Newtonian corrections only to the phase of the signal 
$\phi(t) = 2\,\phi_{\rm orb}(t)$. In this approximation Eq.~(\ref{hpc}) reads:
\begin{subequations}
\ba
h_+ & = & 2 \frac{\Mc^{5/3}}{D_L}\,
\left[ 1 + \left({\bf \hat L} \cdot {\bf \hat N}\right)^2\right]
\left(\pi\,f\right)^{2/3}\,\cos\phi(t)\,,
\label{hplus}
\\
h_{\times} & = & - 4 \frac{\Mc^{5/3}}{D_L}\,
\left({\bf \hat L} \cdot {\bf \hat N}\right)
\left(\pi\,f\right)^{2/3}
\sin \phi(t)\,,
\label{hcross}
\ea
\end{subequations}
where $D_L$ is the source luminosity distance \cite{Weinberg},
${\bf \hat N}$ the unit vector pointing toward the
binary centre of mass (in our convention the GW propagation direction is 
$-{\bf \hat N}$) and ${\bf \hat L}$ the 
orbital angular momentum unit vector.

We review now the expression for the GW phase $\phi(t)$. The signal  
frequency $f$, twice the orbital one, evolves according to~\cite{BDIWW95} 
\ba
\frac{df}{dt} & = & \frac{96}{5}\, \pi^{8/3}\,\Mc^{5/3}\,f^{11/3}
\biggl [ 1-\left( \frac{743}{336}
+\frac{11}{4} \eta \right) (\pi M f)^{2/3}
+\, (4 \pi - \beta) (\pi M f ) \nonumber \\
& & + \left ( \frac{34103}{18144} + \frac{13661}{2016} \eta
+\frac{59}{18} \eta^2 + \sigma\right )
(\pi M f )^{4/3} \biggr ] \,,
\label{fdot2pn}
\ea
where Eq. (\ref{fdot2pn}) is valid through the post$^2$-Newtonian order
and
\ba
\beta & = & \frac{1}{12}\sum_{i=1}^2
\left[113\left( \frac{m_i}{M}\right)^2+75\eta\right]\,
\left({\bf \hat L} \cdot \frac{{\bf S}_i}{m_i^2}\right)\,,
\label{beta}
\\
\sigma & = & \frac{\eta}{48}\left\{-247 
\left(\frac{{\bf S}_1}{m_1^2} \cdot \frac{{\bf S}_2}{m_2^2}\right)
+ 721 
\left[{\bf \hat L} \cdot \frac{{\bf S}_1}{m_1^2}\right]\,
\left[{\bf \hat L} 
\cdot \frac{{\bf S}_1}{m_1^2}\right]
\right\}
\label{sigma}
\ea
are the so called {\it spin-orbit} 
and {\it spin-spin parameters}, respectively \cite{BDIWW95}.
From Eq. (\ref{fdot2pn}) and $d\phi/dt = 2\pi f$, one can derive the time 
and phase evolution of the gravitational radiation:
\ba
t(f) & = & t_c - 5 \,(8\pi f)^{-8/3}\, \Mc^{-5/3}
\Biggl[ 1
+ \frac{4}{3} \left( \frac{743}{336} +\frac{11}{4} \eta \right)
\,(\pi M f)^{2/3} 
- \frac{8}{5}\,(4\pi - \beta)\,(\pi M f)\nonumber\\
& & \quad +\, 2 \left(\frac{3058673}{1016064} + \frac{5429}{1008}\,\eta +
\frac{617}{144}\,\eta^2 - \sigma\right)\,(\pi M f)^{4/3}\Biggr]\,,
\label{tpn2}
\\
\phi(f) & = & \phi_c - \frac{1}{16}\,(\pi f\Mc)^{-5/3}
\Biggl[ 1
+ \frac{5}{3} \left( \frac{743}{336} +\frac{11}{4} \eta \right)
\,(\pi M f)^{2/3} 
- \frac{5}{2}\,(4\pi - \beta)\,(\pi M f)\nonumber\\
& & \quad +\, 5 \left(\frac{3058673}{1016064} + \frac{5429}{1008}\,\eta +
\frac{617}{144}\,\eta^2 - \sigma\right)\,(\pi M f)^{4/3}\Biggr]\,;
\label{phipn2}
\ea
$t_c$ and $\phi_c$, the {\it time} and {\it phase  at coalescence}, 
are constants of integration, defined as the value
of $t$ and $\phi$ at (formally) $f = \infty$. As we mentioned, massive black holes
spend months-to-years in the LISA observational window, depending on the mass. For
future reference, if we assume that LISA observes the final year of in-spiral of
a binary, then the signal sweeps the frequency band between
\be
f_{\rm a} \simeq 4.2 \times 10^{-5} \left(\frac{T_{\rm obs}}{1\,{\rm yr}}\right)^{-3/8}\,
\left[\frac{\Mc\,(1 + z)}{10^6\,\Ms}\right]^{-5/8}\,,
\label{fa}
\ee
which we call "arrival" (or "initial") frequency, and $f_{\rm isco}$.
An interesting quantity to consider is the total number of gravitational wave cycles 
recorded by the detector:
\be
{\cal N}(t) = \frac{1}{2\pi}\,\int_{t_a}^t \phi(t')\,dt'\,.
\label{Ngw}
\ee
From Eq.~(\ref{phipn2}) and~(\ref{fa}), it is easy to compute the number
of cycles in the final year of in-spiral:
\ba
{\cal N} & \approx & 4.8\times 10^2\, 
\left(\frac{f_{\rm a}}{10^{-4}\,{\rm Hz}}\right)^{-5/3}\,
\left[\frac{\Mc\,(1 + z)}{10^6\,\Ms}\right]^{-5/3}
\nonumber\\
 & \approx & 2.1\times 10^3\,\left(\frac{T_{\rm obs}}{1\,{\rm yr}}\right)^{5/8}\,
\left[\frac{\Mc\,(1 + z)}{10^6\,\Ms}\right]^{-5/8}\,.
\label{Ngw1}
\ea
Table~\ref{tab:wcy} contains reference values for $f_{\rm a}$, 
$f_{\rm isco}$ and ${\cal N}$ for a number of choices of masses and observation times.

\begin{table}[htb!]
\caption{\label{tab:wcy}
Summary of the frequency band swept by massive black hole binary systems 
and the number of wave cycles recorded at the detector output in LISA
observations for selected masses and integration times. 
The table shows the initial and final frequency of the GW's, 
$f_{\rm a}$ and $f_{\rm isco}$ respectively, the corresponding time of
observation $T_{\rm obs}$, and the number of waves cycles recored at the output 
of the detector, coming
from the various contributions of the terms
in Eq.~\protect{(\ref{phipn2})}:
Newtonian: the first term; 
post$^1$-Newtonian (1PN): the second term, proportional to $(\pi\,M\,f)^{2/3}$;
tail: $-10\,\pi\,(\pi\,M\,f)$;
spin-orbit: $(5/2)\,\beta\,(\pi\,M\,f)$; post$^2$-Newtonian (2PN): the last term,
proportional to $(\pi\,M\,f)^{4/3}$ with $\sigma = 0$;
spin-spin: the $-5\,\sigma\,(\pi\,M\,f)^{4/3}$ term. The sources are selected 
with masses in the range $10^7\,\Ms$ - $10^5\,\Ms$;
for each mass choice, three different $f_{\rm a}$'s are considered: the first
corresponds to $T_{\rm obs} = $1 yr, which does depend on $m_1$ and $m_2$,
whereas the other two are the same for all $m_1$ and $m_2$, and
correspond to $f_{\rm a} = 5\times 10^{-5}$ Hz ,
and $10^{-4}$ Hz, respectively.
Notice that depending on $f_{\rm a}$, $m_1$ and $m_2$, systems
are observable for different times. The fiducial source is at $z=1$.
}
\begin{tabular}{|cc|cc|c|cccccc|}
\hline
&&&&& \multicolumn{6}{c|}{number of wave cycles}\\
%\cline{6-11} \\
$m_1$ & $m_2$ & $f_a$ & $f_{\rm isco}$ & $T_{\rm obs}$  & 
Newt. & 1PN & tail & spin-orbit & 
2PN & spin-spin \\
$(\Ms)$ & $(\Ms)$ & ($\times 10^{-4}\,{\rm Hz}$) &  ($\times 10^{-4}\,{\rm Hz}$) &  
(yr) & & & & $(\times \beta)$ & & $(\times \sigma)$ \\
&&&&&&&&&&\\
\hline
$10^         7 $  & $10^         7 $  &                0.070 &                1.099 &                1.000 &             342 &              42 &             -40 &               3 &               3 &              -1\cr
$10^         7 $  & $10^         7 $  &                0.500 &                1.099 &                0.004 &              10 &               3 &              -5 &               0 &               1 &               0\cr
%$10^         7 $  & $10^         7 $  &                1.000 &                1.099 &                0.000 &               1 &               0 &               0 &               0 &               0 &               0\cr
$10^         7 $  & $10^         6 $  &                0.153 &                1.999 &                1.000 &             761 &              87 &            -103 &               8 &               7 &              -2\cr
$10^         7 $  & $10^         6 $  &                0.500 &                1.999 &                0.038 &              96 &              22 &             -34 &               3 &               3 &              -1\cr
$10^         7 $  & $10^         6 $  &                1.000 &                1.999 &                0.005 &              23 &               7 &             -13 &               1 &               1 &               0\cr
$10^         7 $  & $10^         5 $  &                0.350 &                2.177 &                1.000 &            1792 &             290 &            -455 &              36 &              32 &             -10\cr
$10^         7 $  & $10^         5 $  &                0.500 &                2.177 &                0.366 &             949 &             187 &            -319 &              25 &              24 &              -8\cr
$10^         7 $  & $10^         5 $  &                1.000 &                2.177 &                0.045 &             238 &              65 &            -130 &              10 &              11 &              -4\cr
$10^         7 $  & $10^         4 $  &                0.776 &                2.196 &                1.000 &            4102 &             990 &           -1881 &             150 &             153 &             -51\cr
$10^         7 $  & $10^         4 $  &                0.500 &                2.196 &                3.649 &            9477 &            1834 &           -3160 &             252 &             235 &             -78\cr
$10^         7 $  & $10^         4 $  &                1.000 &                2.196 &                0.447 &            2383 &             647 &           -1296 &             103 &             110 &             -37\cr
$10^         6 $  & $10^         6 $  &                0.295 &               10.993 &                1.000 &            1453 &             102 &             -76 &               6 &               5 &              -1\cr
$10^         6 $  & $10^         6 $  &                0.500 &               10.993 &                0.246 &             602 &              59 &             -51 &               4 &               4 &              -1\cr
$10^         6 $  & $10^         6 $  &                1.000 &               10.993 &                0.038 &             187 &              28 &             -30 &               2 &               3 &              -1\cr
$10^         6 $  & $10^         5 $  &                0.647 &               19.987 &                1.000 &            3216 &             215 &            -201 &              16 &              11 &              -3\cr
$10^         6 $  & $10^         5 $  &                0.500 &               19.987 &                1.991 &            4950 &             280 &            -243 &              19 &              12 &              -4\cr
$10^         6 $  & $10^         5 $  &                1.000 &               19.987 &                0.312 &            1552 &             137 &            -144 &              11 &               9 &              -3\cr
$10^         6 $  & $10^         4 $  &                1.508 &               21.768 &                1.000 &            7571 &             748 &            -943 &              75 &              55 &             -18\cr
$10^         6 $  & $10^         4 $  &                0.500 &               21.768 &               19.285 &           48121 &            2366 &           -2176 &             173 &              96 &             -31\cr
$10^         6 $  & $10^         4 $  &                1.000 &               21.768 &                3.019 &           15096 &            1155 &           -1300 &             103 &              68 &             -22\cr
$10^         5 $  & $10^         5 $  &                1.242 &              109.929 &                1.000 &            6161 &             247 &            -141 &              11 &               8 &              -2\cr
$10^         5 $  & $10^         5 $  &                0.500 &              109.929 &               11.261 &           28097 &             618 &            -266 &              21 &              11 &              -2\cr
$10^         5 $  & $10^         5 $  &                1.000 &              109.929 &                1.781 &            8848 &             308 &            -165 &              13 &               9 &              -2\cr
$10^         5 $  & $10^         4 $  &                2.729 &              199.872 &                1.000 &           13594 &             519 &            -374 &              30 &              16 &              -5\cr
$10^         5 $  & $10^         4 $  &                0.500 &              199.872 &               91.802 &          230221 &            2867 &           -1208 &              96 &              33 &              -9\cr
$10^         5 $  & $10^         4 $  &                1.000 &              199.872 &               14.494 &           72508 &            1430 &            -753 &              60 &              25 &              -7\cr
\hline
\end{tabular}
\end{table}

\subsection{Spin induced precession}
\label{subsec:spin}

Present observations and theoretical prejudices suggest that massive 
black holes could be rapidly rotating. As we have mentioned in the 
previous section, if black holes are spinning then the GW phase~(\ref{phipn2}) 
contains two additional terms, one at the 1.5PN order, which is
proportional to
$\beta$, and the other at the 2PN order, which is proportional 
to $\sigma$. However spins introduce a much more dramatic effect in the
structure of the waveform, which is qualitatively different from the 
ones just mentioned: the orbital angular momentum ${\bf L}$ 
and the spins ${\bf S}_1$ and ${\bf S_2}$ change orientation due
to spin-orbit interaction (1.5PN effect), so that the radiation detected has a time varying polarisation. 
${\bf L}$, ${\bf S}_1$ and ${\bf S_2}$ 
precess, over a time scale longer than the orbital period 
($2/f \sim 30$ min) but shorter than the observation time ($\sim 1$ yr),
around the (almost) fixed direction of the total angular momentum
${\bf L} + {\bf S}_1 + {\bf S_2}$. This effect is actually quite 
dramatic and drastically changes the signature of the detected signal, 
as the orientation of the orbital plane changes during the year long
observation time, and the angular momenta complete several precession cycles
around the fixed detection.

The spin-orbit induced precession is the key new physical effect which 
is introduced in this paper. Here we review the main concepts and 
expressions and refer the reader to~\cite{ACST94,Kidder95} for more
details. In this paper we restrict our attention to
the so-called {\it simple precession} \cite{ACST94}, which takes place 
when 
(i) the BH masses are equal ($m_1 \sim m_2$), {\it or} 
(ii) one of the BH's has negligible spin (for convention, in this paper we shall assume
$S_2 = 0$).
Furthermore, one needs to impose the condition that the angular momenta are
oriented so that ${\bf L} \ne - ({\bf S_1} + {\bf S_2})$. Under this
assumptions the equations that describe the evolution of the angular momenta
simplify considerably, and one can construct explicit analytical solutions 
for the relevant quantities. Moreover the regime that 
we consider is well justified form an astrophysical point of view.

For circular orbits through the post$^{1.5}$-Newtonian order the equations describing
the evolution of ${\bf L}$, ${\bf S}_1$ and ${\bf S_2}$ read
\be
{\bf{\dot {\hat L }}} = \left[\frac{1}{r^3} 
\left(2+ \frac{3m_2}{2m_1}\right) 
{\bf J}\right] \times {\bf \hat L}\,,
\ee
\be
\dot{{\bf \hat S}} = \left[\frac{1}{a^3}
\left(2+ \frac{3m_2}{2m_1}\right) 
{\bf J}\right] \times {\bf \hat S}\,,
\ee
\be
\frac{d}{dt}\left({\bf S}_1 \cdot {\bf S}_2\right) = 0\,,
\ee
\be
\dot{S} = 0\,,
\ee
where 
\be
{\bf J} = {\bf L} + {\bf S}\quad\quad  ({\bf S} = {\bf S_1} + {\bf S_2})
\label{J}
\ee
is the conserved total angular momentum \cite{BO75,TH} 
(indeed ${\dot {\bf L}} = - {\dot {\bf S}}$). 

 Following~\cite{ACST94}, we highlight the main features
of simple precession and provide, in a ready-to-use form,
the relevant quantities needed for the computation of the LISA 
detector output. 
We will neglect the post$^{2}$-Newtonian corrections, 
while retaining the post$^{1.5}$-Newtonian 
terms (spin-orbit), which are dominant, for the $m_1 = m_2$ case; they
vanish anyway for $S_2 = 0$~\cite{ACST94,Kidder95}.

One can summarise the main features of the {\it simple precession} as follows.
${\bf \hat J}$, ${\bf \hat L}$ and ${\bf \hat S}$ precess with 
the same angular velocity
\be
\Omega_p = \left(2+ \frac{3m_2}{2m_1}\right) \frac{J}{r^3}
\label{omegap}
\ee
around the fixed direction
\be
{\bf \hat J}_0 = {\bf \hat J} -\epsilon {\bf \hat J} \times {\bf \hat L}\,,
\label{J0}
\ee
where
\be
\epsilon \equiv \frac{L}{J}\frac{\dot{L}}{L\Omega_p}\,\ll 1.
\label{epsilon}
\ee
The quantities 
${\bf S}_1 \cdot {\bf S}_2$, $S = |{\bf S}_1 + {\bf S}_2|$ and 
\be
\kappa \equiv {\bf \hat S} \cdot {\bf \hat L}
\label{Sigma}
\ee
are constant during the in-spiral. The phase of the gravitational
signal contains an additional contribution, the so-called 
"Thomas precession phase", given by~\cite{ACST94}
\be
\delta_p\phi (t) = -\int_t^{t_c} \delta_p\dot{\phi}(t')\,dt'\,,
\label{dphipr}
\ee
where
\be
\delta_p\dot{\phi}(t) =
\frac{\gras{\hat{L}}\cdot \gras{\hat{N}}}{1-(\gras{\hat{L}}\cdot 
\gras{\hat{N}})^2}
\,(\gras{\hat{L}}\times \gras{\hat{N}})\cdot {\gras{\dot{\hat L}}}\,.
\label{ddphiprdt}
\ee

In analogy with the number of wave cycles, we can introduce here the 
number of precession cycles ${\cal N}_p$ 
that ${\bf \hat L}$ and ${\bf \hat S}$ undergo around ${\bf \hat J}$ 
from some initial frequency $f_a$, at time $t_{\rm a}$, to the final coalescence at $t_c$:
\be
{\cal N}_p = \frac{1}{2\,\pi}\int_{t_a}^{t_c}\,\Omega_p(t)\,dt\,.
\label{Np}
\ee
Depending on the mass ratio of a binary system, whether $m_1 \sim m_2$ ($L \gg S$)
or $m_1 \gg m_2$ ($L \ll S$), ${\cal N}_p$ is well approximated
by the following expressions:
\be
{\cal N}_p \approx
\left\{
\begin{array}{ll}
10 \,
\left(1 + \frac{3\, m_2}{4\,m_1}\right)\,
\left(\frac{m}{10^6\,\Ms}\right)^{-1}\,\left(\frac{f_a}{10^{-4}\,{\rm Hz}}\right)^{-1}
&  \quad \quad (L \gg S)
\\
2\,
\left(1 + \frac{3\, m_2}{4\,m_1}\right)\,\left(\frac{m_1}{m_2}\right)\,
\left(\frac{S}{m_1^2}\right)\,
\,\left(\frac{m}{10^6\,\Ms}\right)^{-2/3}\,\left(\frac{f_a}{10^{-4}\,{\rm Hz}}\right)^{-2/3}
& \quad \quad (L \ll S)
\end{array}\,.
\right.
\label{Np1}
\ee
Notice that the precession cycles accumulate at low frequency, and,  
for $L\gg S$, ${\cal N}_p$ does not depend on $S$; if $L\ll S$,  
${\cal N}_p\propto (m_1/m_2)\gg 1$ which could therefore generate a very large
number of precessions. In Table~\ref{tab:Nprec} we show 
the number of precession
cycles ${\cal N}_p$ and the value of the precession frequency $\Omega_p$ 
at $f_a$ and  $f_{\rm isco}$ for a number of choices of the mass of a binary system.

\begin{table}
\caption{\label{tab:Nprec} The total number of precession cycles ${\cal N}_p$ and the precession angular frequency $\Omega_p$ at the beginning
and end of LISA observations, for the final year of in-spiral 
of a binary system at redshift $z = 1$ with 
$\kappa  = {\bf \hat S} \cdot {\bf \hat L} = 0.9$ for selected values of the masses and spin magnitude.
}
\begin{tabular}{|c|c|c|c|c|c|}
\hline
$m_1$ & $m_2$ & $S/M^2$ & $\Omega_p$ at $f_a$ & $\Omega_p$ at $f_{\rm isco}$ &
Number of precession cycles \\
$(\Ms)$ & $(\Ms)$ & & ($\times 10^{-4}\,{\rm Hz}$) &  ($\times 10^{-4}\,{\rm Hz}$) & \\
\hline
$10^         7 $  & $10^         7 $  &       0.95 &      0.009 &      1.254 &              11\cr
$10^         7 $  & $10^         7 $  &       0.50 &      0.007 &      0.892 &               9\cr
$10^         7 $  & $10^         7 $  &       0.10 &      0.005 &      0.579 &               7\cr
$10^         7 $  & $10^         6 $  &       0.95 &      0.008 &      1.043 &              11\cr
$10^         7 $  & $10^         6 $  &       0.50 &      0.005 &      0.632 &               7\cr
$10^         7 $  & $10^         6 $  &       0.10 &      0.003 &      0.272 &               4\cr
$10^         7 $  & $10^         5 $  &       0.95 &      0.026 &      0.911 &              34\cr
$10^         7 $  & $10^         5 $  &       0.50 &      0.014 &      0.489 &              19\cr
$10^         7 $  & $10^         5 $  &       0.10 &      0.003 &      0.114 &               5\cr
$10^         7 $  & $10^         4 $  &       0.95 &      0.133 &      0.895 &             127\cr
$10^         7 $  & $10^         4 $  &       0.50 &      0.064 &      0.472 &              71\cr
$10^         7 $  & $10^         4 $  &       0.10 &      0.012 &      0.096 &              15\cr
$10^         6 $  & $10^         6 $  &       0.95 &      0.018 &     12.540 &              25\cr
$10^         6 $  & $10^         6 $  &       0.50 &      0.015 &      8.919 &              20\cr
$10^         6 $  & $10^         6 $  &       0.10 &      0.013 &      5.788 &              16\cr
$10^         6 $  & $10^         5 $  &       0.95 &      0.015 &     10.432 &              23\cr
$10^         6 $  & $10^         5 $  &       0.50 &      0.011 &      6.319 &              16\cr
$10^         6 $  & $10^         5 $  &       0.10 &      0.007 &      2.716 &               9\cr
$10^         6 $  & $10^         4 $  &       0.95 &      0.047 &      9.110 &              74\cr
$10^         6 $  & $10^         4 $  &       0.50 &      0.025 &      4.892 &              40\cr
$10^         6 $  & $10^         4 $  &       0.10 &      0.007 &      1.144 &              11\cr
$10^         5 $  & $10^         5 $  &       0.95 &      0.038 &    125.396 &              54\cr
$10^         5 $  & $10^         5 $  &       0.50 &      0.034 &     89.194 &              46\cr
$10^         5 $  & $10^         5 $  &       0.10 &      0.030 &     57.877 &              39\cr
$10^         5 $  & $10^         4 $  &       0.95 &      0.030 &    104.315 &              48\cr
$10^         5 $  & $10^         4 $  &       0.50 &      0.023 &     63.186 &              34\cr
$10^         5 $  & $10^         4 $  &       0.10 &      0.016 &     27.159 &              22\cr
\hline
\end{tabular}
\end{table}

Using Eq.~(\ref{omegap}) we define now the {\it precession angle} $\alpha$ as 
\be
\frac{d\alpha}{dt} = \Omega_p\,,
\label{dalphadt}
\ee
From Eqs.~(\ref{J}), (\ref{J0}) and~(\ref{Sigma}) it is straightforward
to derive the relevant equations for the evolution of the 
total and orbital angular momentum:
\begin{subequations}
\ba
{\bf \hat J} & = & \frac{{\bf \hat L} + \Upsilon {\bf \hat S}}
{\left(1 + 2 \kappa\, \Upsilon + \Upsilon^2\right)^{1/2}}\,,
\label{hatJ}
\\
J & = & L \left(1 + 2 \, \Upsilon + \Upsilon^2\right)^{1/2}\,,
\label{modJ}
\\
{\bf {\dot{\hat J}}} & = & \dot{\Upsilon} \frac{{\bf \hat S} 
( 1 + \Sigma \Upsilon) - 
{\bf \hat L} (\kappa - \Upsilon)}
{\left(1 + 2 \kappa \, \Upsilon + \Upsilon^2\right)^{3/2}}\,,
\label{dothatJ}
\\
{\bf {\dot{\hat L}}}  & = & \Omega_p\left[{\bf \hat J}_0\times {\bf \hat L}
+ \epsilon ({\bf \hat J}_0 \times {\bf \hat L}) \times {\bf \hat L}\right]\,,
\label{dothatL1}
\\
{\bf \hat L} &  = & {\bf \hat J}_0 \cos\lambda_L + 
\frac{{\bf \hat z} - {\bf \hat J}_0 \cos\theta_J}{\sin\theta_J}
\sin\lambda_L \cos\alpha +
({\bf \hat J}_0 \times {\bf \hat z})
\frac{\sin\lambda_L \sin\alpha}{\sin\theta_J}\,.
\label{Lprec}
\ea
\end{subequations}
In the previous equation $\theta_J$ and $\phi_J$ are the polar coordinates
of ${\bf \hat J}_0$ in the fixed frame and
\be
\lambda_L \equiv {\rm arcsin} \frac{|\dot{{\bf \hat L}}|}{\Omega_p}\,
\label{sinlambdaLJ}
\ee
is the angle between ${\bf \hat L}$ and ${\bf \hat J}$ (at the first order in $\epsilon$,
the approximation in which we are working, $\lambda_L$ coincides with the  angle 
between ${\bf \hat L}$ and ${\bf \hat J}_0$) and 
\be
\Upsilon(t) \equiv \frac{S}{L(t)}\,.
\label{U}
\ee
From Eqs. (\ref{omegap}) and (\ref{epsilon}) one can write 
explicitly:
\be
\Omega_p =  \left(2 + \frac{3 m_2}{2 m_1}\right)
\frac{L}{r^3} {\cal G}\,,
\label{omegap1}
\ee
\be
\epsilon =  
\frac{16}{5}\left(\frac{m}{r}\right)^{3/2}\,
\left[\left(1 + \frac{3 m_2}{4 m_1}\right)\,{\cal G}^2
\right]^{-1}\,,
\label{epsilon1}
\ee
where
\be
{\cal G}^2(f) \equiv 
1 + 2 \kappa \Upsilon + 
\Upsilon^2 \,.
\label{G}
\ee
Eqs. (\ref{U}), (\ref{omegap1}), (\ref{epsilon1}) and (\ref{G}) 
can be used to integrate Eq. (\ref{dalphadt}):
\ba
\alpha & = & \alpha_c
- \frac{96}{5 \mu^3 m^3} \left(1 + \frac{3 m_2}{4 m_1}\right)
\Biggl[2\, ({\cal G}\, L)^3 - 3 \kappa\,S \left(L + \kappa\,S\right)\,{\cal G}\, L
\nonumber\\
& & 
- 3\, \kappa\,S^3 \left(1 - \kappa^2\right)
 \, {\rm arcsinh}\left(\frac{ L + \kappa\,S}
{S \left(1 - \kappa^2\right)^{1/2}}\right)\Biggr]\,.
\label{alpha}
\ea
In Eq.~(\ref{alpha}) the angle $\alpha_c$ is a constant of integration that
essentially identifies the position of ${\bf \hat L}$ and ${\bf \hat S}$ on
the precession cone at the reference time $t_c$.
Finally, the frequency evolution of $\lambda_L$ -- the angle between 
${\bf \hat L}$ and ${\bf \hat J}$ changes with time -- is given by:
\begin{subequations}
\ba
\sin\lambda_L & = & \frac{\Upsilon\left(1 - \kappa^2\right)^{1/2}}
{\left(1 + 2 \kappa \, \Upsilon + \Upsilon^2\right)^{1/2}} = 
\frac{S\, \left(1 - \kappa^2\right)^{1/2}}{L(f) {\cal G}(f)}
\,,
\label{sinlL}
\\
\cos\lambda_L & = & \frac{1 + \kappa\,\Upsilon}
{\left(1 + 2 \kappa \, \Upsilon + \Upsilon^2\right)^{1/2}} = 
\frac{L(f) + \kappa\,S}{L(f) {\cal G}(f)}\,.
\label{coslL}
\ea
\end{subequations}

Notice that at the 1.5PN order, which corresponds to the 
the post-Newtonian order at which 
we model the waveform in this paper, the contribution
of the spins to the GW phase~(\ref{phipn2}) depends only on the parameter $\beta$. 
However, if one takes into account the change of orientation of the angular momenta,
then the spin-orbit coupling is described by three 
parameters, say $\alpha_c$, $\kappa$ and $S$, on which $\beta$ depends.  
In the implementation of the computation of the Fisher information matrix, cf
Sec.~\ref{sec:error}, we found convenient to use as independent parameters 
$\beta$, $\kappa$ and $\alpha_c$.

\section{The observed signal}
\label{sec:output}

We can now derive ready-to-use analytical expressions for
the signal measured by LISA at the Michelson detector output, Eq.~(\ref{resp}),
when the source is an in-spiralling binary system of spinning massive
black holes. In this case, due
to the relativistic spin-orbit coupling, the binary angular momenta,
${\bf L}$, ${\bf S}_1$ and ${\bf S}_2$, precess, following the equations
described in Sec.\ref{subsec:spin}. It is worth spelling out the
assumption under which the signal is derived: (i) the binary is in circular
orbit; (ii) the waveform is modelled using the restricted post$^{1.5}$-Newtonian
approximation (amplitude at the lowest quadrupole Newtonian contribution,
and GW phase at the 1.5PN order); (iii) the masses and/or spins are such that the
system undergoes "simple precession".

In the most general case the signal depends on 17 parameters: 
2 mass parameters, 6 parameters related to 
the BH spins -- the magnitude and orientation of each spin --
the orbital eccentricity, the luminosity distance (which is a {\em direct observable}
in GW astronomy), 2 angles identifying the location of the source in the sky, two
angles that describe the orientation of the orbital plane, one angle that describes
the orientation of the ellipse in the orbital plane, an arbitrary reference time, 
say the time at coalescence, and the signal phase at that time. In the
approximation that we consider here the signal depends on 12 unknown 
parameters (cf the next section for more details). The key new 
qualitative feature that we introduce here and that is not 
present in any of the previous analysis~\cite{Cutler98,MH02,Hughes02,SV00} 
is the change of orientation of the orbital angular momentum during the 
observation. Unlike the model considered in~\cite{MH02,SV00}, 
here we consider only the waves emitted at twice the orbital frequency. 

Before providing the explicit expression for $h(t)$, it is worth
summarising the key motions and the associated time scales that affect the detector output:
\begin{itemize}
\item LISA orbits around the Sun, so that the barycentre of the instrument
changes position over the timescale of one sidereal year: this motion Doppler-shifts 
the incoming gravitational waves, and depends only on the source position in the sky; 
\item The orientation of the LISA arms changes over the same time scale
because the detector precesses around the normal to the Ecliptic:
this introduces a phase and amplitude modulation which is due to the
time-dependent response of the detector to the two orthogonal polarisations
$h_+$ and $h_\times$; this effect depends on both the position and orientation
of the source in the sky.
\item The binary orbital plane changes orientation in the sky with ${\bf L}$
and ${\bf S}$ that precess around the
total angular momentum ${\bf J}$, whose direction is essentially constant. The
time scale of precession is longer than the orbital period but much shorter
than the LISA rotation time-scale and depends on both the source-detector relative
orientation and the binary physical parameters
(masses and spins): 
the incoming GW's are therefore {\em intrinsically}
modulated in amplitude and phase. 
\end{itemize}

We consider first the representation of detector output in the
time domain $h^{(\iota)}(t)$; we will then turn to the frequency domain
representation $\tilde h^{(\iota)}(f)$.

\subsection{The measured signal in the time domain}

In the amplitude-and-phase representation, 
the signal $h^{(\iota)}(t)$, cf Eq.~(\ref{resp}), measured by the $\iota$-th Michelson
LISA interferometer and produced by the polarisation amplitudes~(\ref{hplus}) and~(\ref{hcross}) 
reads
\be
h^{(\iota)}(t) = A(t)\, A_p^{(\iota)}(t)\, 
\cos\left[\phi (t) + \delta_p\phi(t) + \varphi_p^{(\iota)}(t) +\varphi_D(t)\right]\,.
\label{meassign}
\ee
In the previous expression
\ba
A_p^{(\iota)}(t) & = & \frac{\sqrt{3}}{2} \left\{
\left[ 1 + \left({\bf \hat L} \cdot {\bf \hat N}\right)^2\right]^2
F_+^{(\iota)}(t)^2 + 4 \left({\bf \hat L} \cdot {\bf \hat N}\right)^2 
F_{\times}^{(\iota)}(t)^2\right\}^{1/2}\,,
\label{polampl}
\\
\varphi_p^{(\iota)}(t) & = & {\rm tan}^{-1}\left\{
\frac{2 \left({\bf \hat L} \cdot {\bf \hat N}\right) F_{\times}^{(\iota)}(t)}
{\left[ 1 + \left({\bf \hat L} \cdot {\bf \hat N}\right)^2\right]
F_+^{(\iota)}(t)}\right\}
\label{polph}
\ea
are the {\it polarisation amplitude} and {\it phase}, respectively;
\be
\varphi_D(t) = 2 \pi R_{\oplus}\, f\,\sin \theta_N \cos(\Phi (t)-\phi_N)
\quad\quad (R_{\oplus} = 1\,{\rm AU})
\label{dopplph}
\ee
is the {\it Doppler phase modulation} induced by the motion of the
LISA detector around the Sun; $\delta_p\phi(t)$ is the so-called 
{\it Thomas precession phase}, Eq.~(\ref{dphipr}),  which vanishes for spin-less
binary systems; the amplitude of 
the gravitational wave signal is
\be
A[t(f)] = 2 \frac{\Mc^{5/3}}{D}\, \left[\pi\,f(t)\right]^{2/3}\,,
\label{Agw}
\ee
and the GW phase $\phi(t)$
is given by Eq.(\ref{phipn2}), where at the 1.5PN order one neglects the term 
proportional to $(\pi M f)^{4/3}$.

%
% FIGURE
%
\begin{figure}
 \includegraphics[height=10cm]{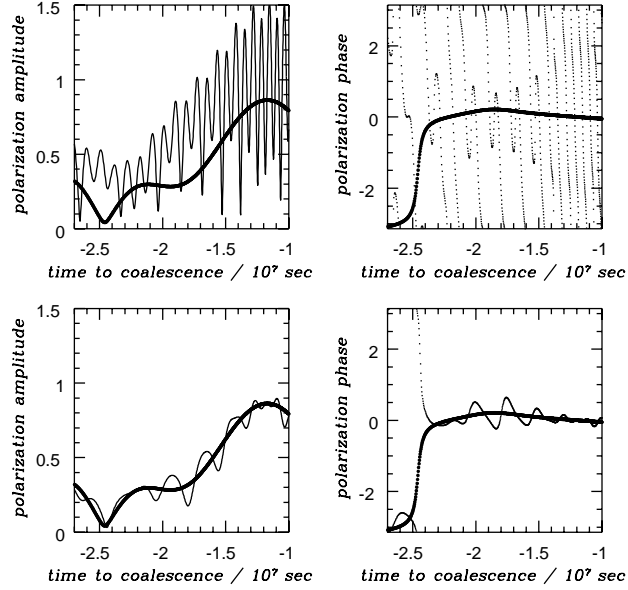}
 \caption{\label{fig:ampphase} 
The evolution of the polarisation amplitude (left panels) and phase (right panels), computed
according to Eqs.~(\ref{polampl}) and~(\ref{polph}),
for the final year of in-spiral of a binary system as a function of time. 
The bold solid-line refers
to black holes with no spin, the dotted-line (for the phase) and thin solid-line (for the
amplitude) to spinning objects. The relevant source parameters are  chosen as follows:
$m_1 = 10^7\,\Ms$, $m_2 = 10^5\,\Ms$, $S/m^2 = 0.95$
and ${\bf \hat S} \cdot {\bf \hat L} =0.5$ (top panels), and
$m_1 = m_2 = 10^6\,\Ms$,  $S/m^2 = 0.3$, and ${\bf \hat S} \cdot {\bf \hat L} =0.9$
(bottom panels). The location and initial orientation of the source have
been chosen randomly. See the text for further details.
}
\end{figure}
%
% END FIGURE
%

The structure of $h^{(\iota)}(t)$ clearly shows that the signal measured at the detector
output is phase and amplitude modulated because of the three effects that
we have mentioned above: 
(ii) the orbital motion of the instrument
around the Sun which Doppler-modulates, through $\varphi_D$, the phase of the
incoming gravitational wave signal; this effect is the same for $h^{(I)}$ and $h^{(II)}$;
(ii) the change of orientation of the detector 
during the time of observation 
that through $F^{(\iota)}_+(t)$ and $F^{(\iota)}_{\times}(t)$ affects
$A_p^{(\iota)}$ and $\varphi_p^{(\iota)}$, in different ways depending on
$\iota$; 
(iii) the change of ${\bf \hat L}$ due to spin-orbit interaction which 
affects $A_p^{(\iota)}$ and $\varphi_p^{(\iota)}$ through $({\bf \hat N} \cdot
{\bf \hat L})$ and $\psi'_N$, cf. Eqs. (\ref{Fplus}), (\ref{Fcross}), 
(\ref{psiNl'}), (\ref{polampl}) and (\ref{polph}); 
the impinging signal is {\it intrinsically} modulated. 
Furthermore $\delta_p\phi$ adds to the phase of the signal.

In order to complete the analysis of the detector output, we need to provide explicit 
expressions for $({\bf \hat L} \cdot {\bf \hat N})$, which enters Eq.~(\ref{polampl})
and~(\ref{polph}), and $({\bf \hat L} \cdot {\bf \hat z}')$ and 
$[{\bf \hat N} \cdot ({\bf \hat L} \times {\bf \hat z}')]$ that are needed to compute
$\psi_N'$, Eq.~(\ref{psiNl'}), and therefore the antenna beam patters. Using Eqs.~(\ref{zprimo})
and (\ref{Lprec}) we obtain:
\ba
{\bf \hat L} \cdot {\bf \hat N} & = &
({\bf \hat J}_0 \cdot {\bf \hat N}) \cos\lambda_L + 
\frac{ \cos\theta_N - ({\bf \hat J}_0 \cdot {\bf \hat N})\cos\theta_J}{\sin\theta_J} 
\sin\lambda_L \cos\alpha
\nonumber\\
& & + \frac{({\bf \hat J}_0 \times {\bf \hat z}) \cdot {\bf \hat N}}{\sin\theta_J}
\sin\lambda_L \sin\alpha\,,
\label{LdotNpr}
\\
{\bf \hat L} \cdot {\bf \hat z'} & = &
({\bf \hat J}_0 \cdot {\bf \hat z}') \cos\lambda_L + 
\frac{ 1 - 2\,({\bf \hat J}_0 \cdot {\bf \hat z}')\cos\theta_J}{2\sin\theta_J} \sin\lambda_L \cos\alpha
\nonumber\\
& & + \frac{({\bf \hat J}_0 \times {\bf \hat z}) \cdot {\bf \hat z}'}{\sin\theta_J}
\sin\lambda_L \sin\alpha\,,
\label{Ldotz'pr}
\\
{\bf \hat N} \cdot ({\bf \hat L} \times {\bf \hat z'}) & = &
{\bf \hat N} \cdot ({\bf \hat J}_0 \times {\bf \hat z'}) \cos\lambda_L + 
\frac{{\bf \hat N} \cdot ({\bf \hat z} \times {\bf \hat z}') - 
{\bf \hat N} \cdot ({\bf \hat J}_0 \times {\bf \hat z}')\cos\theta_J}
{\sin\theta_J} \sin\lambda_L \cos\alpha
\nonumber\\
& & + \frac{{\bf \hat N} \cdot ({\bf \hat J}_0 \times {\bf \hat z}) \times {\bf \hat z}'}
{\sin\theta_J}\sin\lambda_L \sin\alpha\,.
\\
\label{NdotLvectz'pr}
\ea
In the previous expressions $\sin\lambda_L$, $\cos\lambda_L$ and are $\alpha$ are given by 
Eqs.~(\ref{sinlL}),~(\ref{coslL}) and~(\ref{alpha}), respectively, and 
the remaining quantities describing the orientation of the source with respect of the
detector read:
\ba
{\bf \hat J}_0 \cdot {\bf \hat z}' & = & \frac{1}{2} \cos\theta_J - \frac{\sqrt{3}}{2}
\sin\theta_J \cos(\Phi(t) - \phi_J)\,,
\label{thetaJ'}
\\
{\bf \hat J}_0 \cdot {\bf \hat N} & = &
\cos\theta_J \cos\theta_N + \sin\theta_J\sin\theta_N \cos(\phi_J-\phi_N)\,,
\label{JdotNpr}
\\
{\bf \hat{N}} \cdot \left({\bf \hat{J}}_0 \times {\bf \hat{z}}'\right) 
& = &
\frac{1}{2}\sin\theta_N \sin\theta_J \sin ( \phi_J - \phi_N)
\nonumber\\
& & -\frac{\sqrt{3}}{2} \cos\Phi(t) \left(\cos\theta_J \sin\theta_N \sin\phi_N 
- \cos\theta_N \sin\theta_J \sin\phi_J\right)
\nonumber\\
& & - \frac{\sqrt{3}}{2}\sin\Phi(t)\left(
\cos\theta_N \sin\theta_J \cos\phi_J -
\cos\theta_J \sin\theta_N \cos\phi_N\right)\,,
\label{NdotJvectz'pr}
\\
{\bf \hat N} \cdot ({\bf \hat J}_0 \times {\bf \hat z}) & = & 
\sin\theta_N\sin\theta_J \sin(\phi_J - \phi_N)\,,
\label{NdotJvectzp}
\\
{\bf \hat N} \cdot ({\bf \hat z} \times {\bf \hat z}') & = &  \frac{\sqrt{3}}{2}\,
\sin\theta_N\sin\left[\Phi(t) - \phi_N\right]\,,
\label{Ndotzvectzp}
\\
{\bf \hat J} \cdot ({\bf \hat z} \times {\bf \hat z}') & = &  \frac{\sqrt{3}}{2}\,
\sin\theta_J\sin\left[\Phi(t) - \phi_J\right]\,,
\label{Jdotzvectzp}
\\
{\bf \hat N} \cdot ({\bf \hat J}_0 \times {\bf \hat z}) \times {\bf \hat z}' & = &
-\frac{1}{2} \sin\theta_J\left[
\sqrt{3}\cos\theta_N \cos\left[\Phi(t) - \phi_J\right] +
\sin\theta_N \cos(\phi_J-\phi_N) \right]\,.
\label{NdotJvectvectzp}
\ea

Lastly, we need to compute the scalar and vector products necessary to derive the Thomas 
precession phase $\delta_p\phi(t)$, Eqs.~(\ref{dphipr}) and (\ref{ddphiprdt}):
\be
(\gras{\hat{L}} \times \gras{\hat{N}}) \cdot \dot{\gras{\hat{L}}} =
\Omega_p \left[\cos\lambda_L\,
({\bf \hat{N}} \cdot {\bf \hat{L}})
- ({\bf \hat{N}} \cdot {\bf \hat{J}_0})
- \epsilon\, {\bf \hat J}_0 \cdot ({\bf \hat L} \times {\bf \hat N})
\right]\,,
\label{LvectNdotLdotpr}
\ee
\be
{\bf \hat J}_0 \cdot ({\bf \hat L} \times {\bf \hat N}) =
\frac{{\bf \hat N} \cdot ({\bf \hat J}_0 \times {\bf \hat z})}{\sin\theta_J} 
\sin\lambda_L \cos\alpha
+ \frac{\cos\theta_J ({\bf \hat N} \cdot {\bf \hat J}_0) - \cos\theta_N}{\sin\theta_J} 
\sin\lambda_L \sin\alpha\,;
\ee
$\delta_p\phi(t)$ can be therefore computed using Eqs. (\ref{alpha}), 
(\ref{sinlL}), (\ref{coslL}), 
(\ref{LdotNpr}), (\ref{thetaJ'}), (\ref{JdotNpr}) and 
(\ref{NdotJvectz'pr}).

An example of the LISA output when black holes rapidly spin is shown 
in Figure~\ref{fig:ampphase}, where we plot the amplitude and phase modulation, Eqs~(\ref{polampl})
and~(\ref{polph}), during the final year of in-spiral. Two cases are actually presented:
a binary system with 
$m_1 = 10^7\,\Ms$, $m_2 = 10^5\,\Ms$, $S/m^2 = 0.95$
and ${\bf \hat S} \cdot {\bf \hat L} =0.5$ and an equal mass binary where the
parameters correspond to 
$m_1 = m_2 = 10^6\,\Ms$,  $S/m^2 = 0.3$, and ${\bf \hat S} \cdot {\bf \hat L} =0.9$. 
The evolution of $A_{p}(t)$ and $\varphi_p(t)$ for the two sets of
physical parameters is compared to the spin-less case. 
A simple eye inspection shows how dramatic is the effect induced by spins, and 
it should therefore not come as a surprise the fact that the errors associated to
the measurements of the source parameters are strongly affected.

\subsection{The measured signal in the frequency domain}

For several applications in data analysis, it is often more useful to work 
in the frequency domain, and we will derive now an approximation to the
Fourier representation of the signal $h^{(\iota)}(t)$, Eq.(\ref{meassign}).
Our convention for the Fourier transform $\tilde g(f)$ of any 
real function $g(t)$ is:
\be
\tilde g(f) = \int_{-\infty}^{\infty}\, e^{2\pi i f t}\,g(t)\,dt\,.
\label{Fourtr}
\ee

The Fourier transform (\ref{Fourtr}) of the signal $h^{(\iota)}(t)$ can be
computed in an rather straightforward way using the stationary phase
approximation, see~\cite{CF94}, and considering the gravitational waveform $A(t)\cos\phi(t)$ in
Eq.~(\ref{meassign}) as the {\it carrier signal} modulated by the motion of
the detector and the source's orbital plane~\cite{ACST94,Cutler98}. 
Under this assumption, the Fourier transform of the detector output~(\ref{meassign}) reads:
\be
\tilde h^{(\iota)}(f) \simeq
\left\{
\begin{array}{ll}
{\cal A}\,A_p^{(\iota)}[t(f)]\,f^{-7/6}
\,e^{\left\{i\,\left[
\Psi(f) - \varphi_p^{(\iota)}\left[t\left(f\right)\right] -
\varphi_D\left[t\left(f\right)\right] - \delta_p\phi\left[t\left(f\right)\right]\right]\right\}}
\quad\quad & \quad\quad 0 < f \le f_{\rm isco} \\
0 \quad\quad & \quad\quad f > f_{\rm isco}
\end{array}
\right.\,,
\label{hf}
\ee
where
\ba
{\cal A} & = & \left(\frac{5}{96}\right)^{1/2}\, \pi^{-2/3}\,\frac{\Mc^{5/6}}{D}
\label{calA}
\\
\Psi(f) & = & 2\pi f t_c - \phi_c - \frac{\pi}{4} +
\frac{3}{4}\,(8 \pi \Mc f)^{-5/3} 
\, \biggl[1+ \frac{20}{9}\left(\frac{743}{336}+ \frac{11}{4}\,
\eta\right)\,(\pi m f)^{2/3} 
\nonumber\\
& & - 4 \left(4 \pi  - \beta\right)\, (\pi m f) + 
10 \left(\frac{3058673}{1016064} + \frac{5429}{1008}\,\eta +
\frac{617}{144}\,\eta^2 - \sigma\right)\,(\pi m f)^{4/3}
 \biggr]\,.
\label{Psi}
\ea
Notice that in Eq.~(\ref{Psi}) the term proportional to $(\pi m f)^{4/3}$ correspond to the
post$^2$-Newtonian order, and will not be considered in this work.

\section{Parameter estimation}
\label{sec:error}

In this section we discuss the errors associated to the parameter measurement when 
LISA monitors binary systems of rapidly rotating massive black holes in circular orbit.
We start by briefly recalling the general concepts and formulae regarding parameter
estimation -- we refer the reader to~\cite{WZ62,Helstrom68,Finn92,CF94,NV98} and 
references therein for more details -- and then present the results of our analysis 
applied to LISA data.

\subsection{Review of signal analysis and parameter estimation}

The signal $s(t)$ registered at the detector output is the superposition of 
noise $n(t)$ and gravitational waves $h(t;{\gras \lambda})$:
\be
s(t) = h(t;{\gras \lambda}) + n(t)\,;
\ee
${\gras \lambda}$ represents the vector of the unknown parameters (location, masses, spins,
etc.) that characterise the actual waveform and that one wishes to estimate from
the data stream. We assume the noise to be stationary and Gaussian, characterised by
a noise spectral density $S_n(f)$.
In the geometrical approach to signal processing it is useful to consider $s(t)$
as a vector in the signal vector space, and to 
introduce the following inner product between two signals $v$ and $w$~\cite{CF94}:
\be
\left(v|w\right) = 2 \int_0^{\infty} \frac{\tilde v(f) \tilde w^*(f) +
\tilde v^*(f) \tilde w(f)}{S_n(f)}\,df\,.
\label{inner}
\ee
According to the definition~(\ref{inner}), the optimal signal-to-noise ratio (SNR) at which 
$h$ can be detected is
\be
S/N = \frac{(h|h)}{{\rm rms}[(h|n)]} = (h|h)^{1/2}\,.
\label{snr}
\ee
In this paper we discuss the errors associated to the measurement of the unknown
parameter vector ${\gras \lambda}$ that characterises the signal $h(t;{\gras \lambda})$. 
In the limit of large SNR, which is clearly the
case for LISA observations of massive black hole binary systems, the
errors $\Delta {\gras \lambda}$ follow a Gaussian probability distribution:
\be
p(\Delta {\gras \lambda}) = \left(\frac{\det({\bf \Gamma})}{2 \,\pi}\right)^{1/2}
\,e^{-\frac{1}{2}\,\Gamma_{jk} \Delta\lambda^j \Delta\lambda^k}\,.
\label{pl}
\ee
In Eq.(\ref{pl}) the matrix $\Gamma_{jk}$ is known as the Fisher information matrix,
and reads
\be
\Gamma_{jk}^{(\iota)} \equiv 
\left(\frac{\partial h^{(\iota)}}{\partial \lambda^j} \Biggl|\Biggr.
\frac{\partial h^{(\iota)}}{\partial \lambda^k}\right)\,.
\label{fisher}
\ee
The {\it variance-covariance matrix} is simply given by the inverse of the Fisher
information matrix:
\be
\Sigma^{jk} = \left\m \Delta\lambda^j\,\Delta\lambda^k \right\M = 
\left[\left({\bf \Gamma}^{(\iota)}\right)^{-1}\right]^{jk}\,.
\label{vc}
\ee
The matrix ${\gras \Sigma}$ contains full information about the parameter errors and their 
correlations, and is what we need to compute (cf next Section) in order to investigate the LISA
parameter estimation. In fact the diagonal elements of ${\gras \Sigma}$
represent the expected mean squared errors 
\be
\m (\Delta\lambda^j)^2\M = \Sigma^{jj}\,,
\label{mse}
\ee
and its off-diagonal elements provide information about the correlations
among different parameters through the correlation coefficients $c^{jk}$:
\be
c^{jk} = \frac{\Sigma^{jk}}{\sqrt{\Sigma^{jj}\,\Sigma^{kk}}}\quad\quad (-1 \le c^{jk} \le +1)\,.
\label{corr}
\ee
In the limit of high signal-to-noise ratio, $\Sigma^{jj}$ provides a tight
lower bound to the minimum mean-squared error $\m (\Delta\lambda^j)^2\M$ 
the so-called Cramer-Rao bound~\cite{Helstrom68,NV98}.
It is important to notice that the errors~(\ref{mse}) and the correlation coefficients~(\ref{corr})
depend both on the actual value of the signal parameter vector ${\gras \lambda}$. For the case 
of observations with two or more detectors with uncorrelated
noise, the Fisher information matrix is simply:
$\Gamma_{jk} = \sum_{\iota}\,\Gamma_{jk}^{(\iota)}$.

One of the properties that we are interested in is the angular resolution of the instrument,
which we define as:
\be
\Delta \Omega_N  = 2 \pi\,
\left\{
\left\m\Delta\cos\theta_N^2\right\M\,\left\m\Delta\phi_N^2\right\M -
\left\m \Delta\cos\theta_N\,\Delta\phi_N \right\M^2
\right\}^{1/2}\,;
\label{DOmega_N}
\ee
The physical meaning of $\Delta \Omega_N$ is the following: the probability
of the source to lie {\it outside} an
(appropriately shaped) error ellipse enclosing a solid angle 
$\Delta \Omega$ is  simply $e^{-\Delta \Omega/\Delta \Omega_N}$.

\subsection{LISA observations}

We discuss now how accurately LISA can measure the source parameters in the
case of rapidly spinning black holes. 
The goal of this section is to investigate
whether spins, which have been neglected in studies carried out
so far, can significantly affect the estimation of the source parameters. 
A thorough exploration of
this effect requires to probe a very large multi-dimensional parameter
space, and it is well outside of the scope of this paper. 
Here we concentrate on a fiducial source of two $10^6\,\Ms$ black holes at
redshift $z=1$ -- a rather typical LISA source -- and we
compare the errors associated to the parameter measurements in the
case in which spins are present  -- for large spins, $S/M^2 = 0.95$,
and moderate spins $S/M^2 = 0.3$; in both cases $\kappa = 0.9$ --
and the case where spins are neglected (the black holes are considered, a priory, to be
not spinning). 

Before presenting the results we spell out the assumptions under which we
compute the errors~(\ref{mse}):

\begin{itemize}

\item 
The orbit is circular; we regard such hypothesis as 
realistic, as we are dealing 
with massive systems of black holes of comparable mass. As the orbit
shrinks due to dynamical friction the eccentricity is likely to decrease~\cite{BBR80,RR95,VCP94} 
and radiation reaction completes the circularisation process before 
GW's enter the observational window
of LISA; in fact the eccentricity $e$ evolves according to 
$e \propto f^{-19/18}$~\cite{Peters64}.

\item 
The black holes are spinning, so that the binary orbit precesses in
space. However, in oder to simplify the description of the precession motion, still
addressing a realistic astrophysical scenario, we
assume that either $m_1 = m_2$ (the case considered in this paper) 
or $S_2 = 0$ and the spins and angular momentum are
not anti-aligned. Under this conditions a binary undergoes the so-called
simple precession~\cite{ACST94}, and the
equations describing the evolution of the relevant physical quantities simplify
considerably (cf Sec.~\ref{subsec:spin} for more details).

\item 
We restrict the analysis only to the in-spiral phase of the whole coalescence.
We approximate
the waveform $\tilde h^{(\iota)}(f;{\bf \lambda})$ at the restricted 1.5PN
order. The GW signal detected at the LISA output is therefore described 
by Eqs.~(\ref{hf}),~(\ref{calA}), and~(\ref{Psi}).
Because we retain terms up to 1.5PN order in the GW phase, in the expression
of $\Psi(f)$, Eq.~(\ref{Psi}), we neglect the last term proportional to 
$(\pi M f)^{4/3}$. We shut-off the in-spiral waveform at the frequency $f_{\rm isco}$,
given by Eq.~(\ref{fisco}). 
In the computation of the errors~(\ref{mse}) we actually neglect
the Thomas precession phase $\delta_p(t)$, Eq.~(\ref{dphipr}). This 
simplification is motived by computational reasons, and does not affect
in any significant way the final results. In fact, $\delta_p[t(f)]$ 
must be computed numerically at each frequency using the past history 
of the binary; $\delta_p[t(f)]$ needs then to be included in the integrand
of the scalar product(~\ref{inner}) that leads to determination of the elements of
the Fisher information matrix. This double numerical integration, which
needs to be carried out to high accuracy in order to keep under control
numerical instabilities that occur otherwise in the numerical inversion
of $\Gamma_{jk}$, makes the computational time very long, due to our
limited computational resources. We have checked for a few (random) choices
of the source parameters that including $\delta_p[t(f)]$ does not change in
any appreciable way the result. This is to be expected. For the physical parameters
considered in this paper the Thomas precession phase contribute to $\approx 1$
wave cycle, out of a total $\approx 1500$. It also represents a secular increase in 
the phase of the GW signal, qualitatively not different from the one given by $\Psi(f)$.
\item 
We consider the fiducial sources to be at redshift $z=1$~\cite{norm}.
As the systems are at cosmological distance, the values of all the physical 
parameters entering the
GW signal presented in the previous sections must be considered as the 
{\em observed ones}; they differ
from the values of the parameters as measured in the source rest 
frame by a factor $(1+z)$; the parameters are "Doppler-shifted" according to:
\ba
f & \rightarrow & \frac{f}{(1+z)}\,, \nonumber \\
t & \rightarrow & (1+z)\,t\,, \nonumber \\
\Mc & \rightarrow & (1+z)\,\Mc \,,\nonumber \\
m & \rightarrow & (1+z)\,m \,,\nonumber \\
\mu & \rightarrow & (1+z)\,\mu\,.
\ea

\item
We assume that the instrument observes the whole final year of in-spiral;
the GW's sweep therefore the frequency band between the 
{\em arrival frequency} $f_{\rm a}$ and the final cut-off frequency
$f_{\rm isco}$; $f_{\rm a}$ is determined so that after 1 yr,
as measured by an observer in our solar system, the GW
instantaneous frequency reaches $f_{\rm isco}$:
\ba
T_{\rm obs} & = & t(f_{\rm isco}) - t(f_{\rm a})\nonumber\\
& = & 1\,{\rm yr} = 3.1556926\times 10^7\, {\rm sec}
\label{Tobs}
\ea
The range $f_{\rm a} \le f \le f_{\rm isco}$ determines the 
integration domain in~(\ref{fisher}); 
values of $f_{\rm a}$ and $f_{\rm isco}$ for selected
choices of the source parameters are given in Table~\ref{tab:wcy}. Note that we do
not impose a low-frequency cut-off to LISA; this choice is based on the fact that the 
real low frequency noise wall of space-based instruments is not very well understood 
at the moment, placed somewhere in the range $10^{-5}$ Hz - $10^{-4}$ Hz.

\item
The total noise that affects the observations is given by the superposition of
instrumental sources and astrophysical foregrounds of unresolved radiation due to (mainly) 
galactic white dwarf binary systems~\cite{Hilsetal90,BH97},
the so-called "confusion noise". The total noise
spectral density $S_n(f)$ is therefore the sum of these two components, and
we use the analytical approximations given in~\cite{Cutler98,norm} .

\item 
Out of the 17 parameters on which the most general waveform
depends, the signal $\tilde h^{(\iota)}(f;{\bf \lambda})$ that we consider
here depends on 12 independent parameters. In our analysis we adopt the following
choice of independent parameters:
$\ln\Mc$ and $\ln\mu$ (mass parameters), $\cos\theta_N$, $\phi_N$,
$\cos\theta_J$, $\phi_J$, and $\alpha_c$ (geometry of the binary
with respect to the detector), $\kappa$ and $\beta$ (spin parameters),
$\ln D_L$ (distance of the source) and finally $t_c$ and $\phi_c$.

\item 
We compute the expected mean square errors $\left(\m \Delta\lambda_j^2\M\right)^{1/2}$ and
the angular resolution $\Delta\Omega_N$ for the case of observations carried
out with one detector and with combined 
observations of both detectors $I$ and $II$. The analysis is done in the 
the frequency domain: we first compute analytically the derivatives
$\partial h^{(\iota)}/\partial \lambda^j$, where $j = 1,..,12$, then compute 
numerically $\Gamma_{jk}$ and $\Sigma^{jk}$, Eqs.~(\ref{fisher}) and~(\ref{vc});
the integration and matrix inversion are performed using numerical routines of the
NAG library.

\end{itemize}

%
% FIGURE
%
\begin{figure}
 \includegraphics[height=10cm]{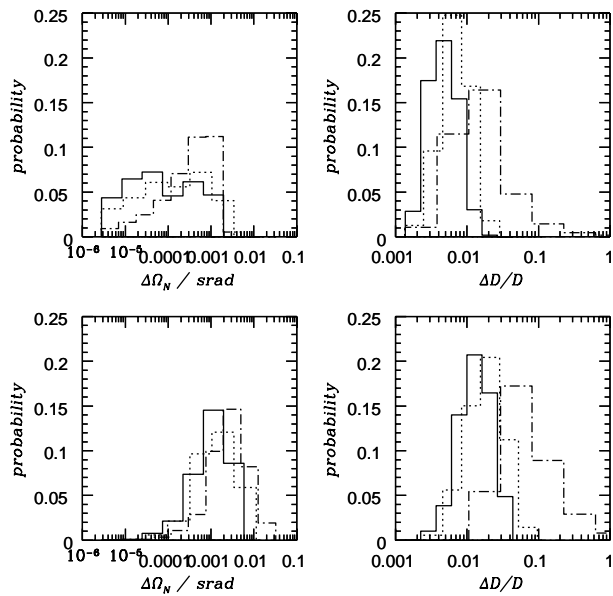}
 \caption{\label{fig:errND} 
The probability distribution of $\Delta\Omega_N$ (left)
and $\Delta D_L/D_L$ (right), for observations of the final year of in-spiral
of a binary system with $m_1 = m_2 = 10^6\,\Ms$ at $z=1$. 
The histograms show the result of a Monte-Carlo simulation, where 1000
sources have  
been randomly located and oriented in the sky.
The bottom panels refer to measurements carried out with only one detector, 
whereas the top panels describe the results obtained by combining the
two independent data streams.
The plots compare the errors for different values of the BH spins:
$S/m^2 = 0.9$ (solid line), 0.3 (dotted line), and 0
(dotted-dashed line).
}
\end{figure}
%
% END FIGURE
%

LISA parameter estimation strongly depends (the results vary by orders of
magnitude) on the actual value of the signal parameters. In particular,
one of the key set of parameters that affect the errors is the location and orientation
of a source with respect to the detector. This represents already a large parameter space 
that one needs to
explore in order to obtain meaningful results. We perform this exploration by means of Monte-Carlo
simulations, which therefore affect the way in which the results are presented,
given in terms of probability distributions. For each set of physical parameters,
we select randomly the 5 geometrical parameters ($\theta_N$, $\phi_N$, $\theta_J$, $\phi_J$
and $\alpha_c$) from a uniform distribution in 
$\cos\theta_N$, $\phi_N$, $\cos\theta_J$, $\phi_J$ and $\alpha_c$. The Monte-Carlo
simulation is done on 1000 different sets of angles. As far as the other 7 parameters
are concerned we chose them as follows. 
We consider a fiducial source at redshift $z = 1$, with $m_1 = m_2 = 10^6\,\Ms$
(which sets the 3 parameters $D_L$, $\Mc$ and $\mu$), with $t_c = \phi_c = 0$. We fix
the {\em tilt angle} parameter $\kappa = 0.9$, and we explore three different values of
the size of the spin $S$: 
(i) $S/M^2 = 0.95$, (ii) $S/M^2  = 0.3$ and (iii) $S = 0$. Notice that 
given $m_1$, $m_2$, $S$ and $\kappa$ we can easily derive $\beta$. 
It is worth noticing that in this paper we do not explore the effect of the tilt
angle $\cos^{-1}\kappa$, which is also likely to affect the errors. Such analysis
is currently in progress~\cite{Vecchioetal03}. We would also like to stress that for the 
case $S = 0$ the waveform that we actually
consider is the one corresponding to the non-precessing restricted 2PN approximation (this is the
waveform used in~\cite{Hughes02}, very similar to the one used in~\cite{Cutler98},
which is the non-precessing restricted 1.5PN approximation). This is equivalent to
assuming that we know {\it a priori} that spins are zero and the binary system does no precess. Note
that this is different from estimating the source parameters using
the waveform~(\ref{hf}) where precession is included, and assuming $S\ll 1$. In the case $S = 0$ 
we need to estimate only 11 parameters, and not 12. The reason of this choice 
is to compare existing results in the literature with our new ones that take into account
spin-orbit modulations.

%
% FIGURE
%
\begin{figure}
 \includegraphics[height=10cm]{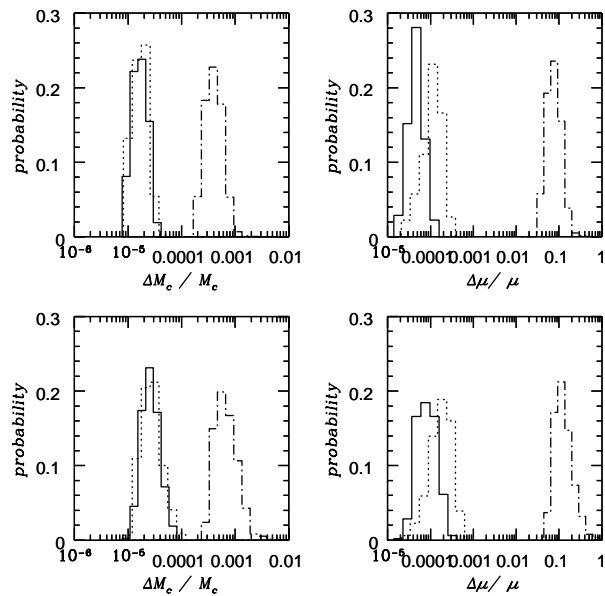}
 \caption{\label{fig:errM} 
The probability distribution of $\Delta\Mc/\Mc$ (left)
and $\Delta\mu/\mu$ (right) for observations carried out with only one
interferometer (bottom) and the two combined LISA outputs (top). The
parameters are the same as in Fig~\ref{fig:errND}.
}
\end{figure}
%
% END FIGURE
%

\begin{figure}
 \includegraphics[height=10cm]{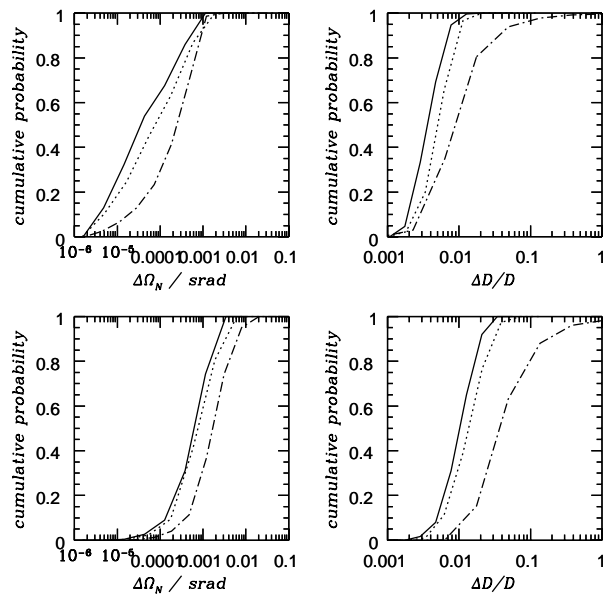}
 \caption{\label{fig:errNDint} 
Same as Figure 2, but now the cumulative probability distributions as a function of
the parameters $\Delta\Omega_N$ and $\Delta D_L/D_L$ are shown. See text for more details
}
\end{figure}

Figs~\ref{fig:errND},~\ref{fig:errM} and~\ref{fig:errNDint} summarise the key
results. Figs~\ref{fig:errND},~\ref{fig:errM} show the probability distribution 
of the parameter errors, for $\Delta\Omega_N$, $\Delta D_L/D_L$,
$\Delta\Mc/\Mc$ and $\Delta\mu/\mu$.
Fig~\ref{fig:errNDint} shows the cumulative probability distributions: if we define
$\xi \equiv \Delta\lambda^j$, the plots show 
$P( \xi < \xi_0) = \int_0^{\xi_0} p(\xi') d\xi'$ as a function of $\xi_0$.
Each plot compares the non-spinning case with the cases in which the black holes are
spinning with two different value of the spins, $S/M^2 = 0.95$ and  0.3 .

The key result, which is absolutely clear from 
Figs~\ref{fig:errND},~\ref{fig:errM} and~\ref{fig:errNDint} is that the errors 
in the spinning-case are smaller than in the non-spinning one, even if the number of
parameters that one needs to estimate is greater (12 instead of 11). One can intuitively
understand this behaviour by looking at Figure~\ref{fig:ampphase}, which shows the
amplitude and phase evolution of the signal recorded at the LISA output: precessing
binaries show a much greater richness of features than non-spinning sources. These
are the features that help in measuring the parameters. 

However, despite all the parameters are estimated with smaller errors with respect
to the non-spinning case, there is striking difference between the "position parameters"
-- location in the sky and luminosity distance -- and mass parameters. 
In fact, the errors $\Delta\Omega_N$ and $\Delta D_L/D_L$ are reduced by a factor
2-to-10 depending on the actual value of the source parameter. We know that LISA
reconstruct the position of the source in the sky by exploiting the modulation in the
amplitude and phase of the signal, and precession introduces one additional modulation effect
which can improve the determination of ${\bf {\hat N}}$ and de-correlates $D_L$ from
$\Mc$, which improves the measurement of the luminosity distance. Still, the change of 
position and orientation of LISA is the main effect that allows the source to be located.
The improvement in $\Delta{\Omega}_N$ and $\Delta D_L/D_L$ is however significant:
the fraction of sources that can be located within, say, one square degree is 10\% for
non-spinning binaries, 35\% for binaries with 0.3 and 50\% with 0.95 (for $\kappa = 0.9$).
The systems whose distance is known to better than 1\% is about 60\% for non-spinning
binaries, and essentially the totality for spinning binaries.

The situation is radically different for the masses: in the non-spinning case
the instrument reconstructs the value of the two mass parameters by staying in
phase with the GW phase, whose time evolution is mainly controlled by $\Mc$ (the
leading Newtonian term) and, to a less extent, by $\mu$, through the post-Newtonian 
corrections,
cf Eqs.~(\ref{phipn2}),~(\ref{Psi}) and Table~\ref{tab:wcy}. $\Mc$ and $\mu$ play no role
in the amplitude and phase modulation, $A_p(t)$ and $\varphi_p(t)$, Eqs~(\ref{polampl})
and~(\ref{polph}). As a
consequence, $\Mc$ is measured much more accurately, by a factor $\sim 100$ or more, than 
$\mu$, see Fig~(\ref{fig:errM}) and~\cite{Cutler98,Hughes02}. If spins are present, 
the masses start to play a role also in $A_p(t)$ and $\varphi_p(t)$,
because they control the rate at which ${\bf L}$ and ${\bf S}$ precess, and therefore 
are responsible for the intrinsic amplitude and phase modulation of the signal. 
This has two effects on
the parameter estimation: it de-correlates $\Mc$ from $\mu$ (at the Newtonian
order the two mass parameters are even degenerate) and provides new features in the measured
signal that 
improve the parameter estimation. The net effect is that the errors in $\Mc$ and $\mu$ are
now of the same order, with a drastic improvement with respect to the non-spinning case,
of by a factor $\sim 50$ and $\sim 10^3$ for $\Delta \Mc/\Mc$ and $\Delta\mu/\mu$, 
respectively;
$\Delta \Mc/\Mc$ is still $\approx 3$ times smaller than $\Delta\mu/\mu$, due to the role
played by $\Mc$ in the GW phase $\phi(t)$. It also clear that the additional signature
produced by the masses on the amplitude and phase modulation helps in removing the
correlation between $\Mc$ and $D_L$, therefore reducing $\Delta D_L/D_L$, as we have
already mentioned.

To summarise, the parameters of two $10^6\,\Ms$ black holes spiralling toward the final 
merger at $z = 1$, and rapidly rotating, say $S/M^2 = 0.95$ and $\kappa = 0.9$, can be 
measured very accurately: $\Mc$  and $\mu$ can be measured to a few parts in $10^5$, 
the luminosity distance to better than $1\%$ (in some cases almost $0.1\%$), and with an
error box in the sky $10^{-6}\,{\rm srad} \simlt \Delta\Omega_N \sim 10^{-3}\,{\rm srad}$.
Notice that the errors scale with the distance roughly as $\Delta\Omega_N \sim 1/D_L^2$
and $\Delta \Mc/\Mc \sim \Delta \mu/\mu \sim \Delta D_L/D_L \sim 1/D_L$. 
This would be strictly
true for white noise, which is not the case for LISA. In particular going to higher redshift,
one can expect a degradation of the measurements more severe than the one predicted
by this simple scaling, because GW's are red-shifted to lower frequencies where
the noise is higher and some initial portion of the signal falls out of the observational
window.

%
% FIGURE
%
%
% END FIGURE
%

For multi-band, electro-magnetic and gravitational, observations of the same event
the key parameter is the instrument's angular resolution. In fact there is some
hope that LISA will have enough angular resolution to locate the galaxy or
galaxy cluster where a coalescence of massive black holes takes place; this would
enable other telescopes to be pointed at the same area of the sky and to observe
the aftermath of such catastrophic event. However, previous 
studies~\cite{Cutler98,MH02,Hughes02,SV00} have concluded that this might actually
be impossible for sources at $z\approx 1$ (or beyond), because of the poor angular resolution
of GW observations. In short, the results of our analysis suggest that in some exceptional case
LISA might be able to identify the host galaxy cluster of a MBH binary of $10^6\Ms$ at $z = 1$, but
in general the LISA error box will contain several hundreds of galaxy clusters and $\simgt 10^5$
galaxies, all of them potential hosts of a LISA detected source. Such conclusion can be easily
derived by comparing the LISA angular resolution with the one of other telescopes and the
angular size of galaxies at high redshift.
An optical telescope, such
as the Keck telescope, has a field of view of about 1 square degree. 
Chandra has a square field of view of $16\times 16$ arcmin with a spatial resolution 
of $\simeq 1$ arcsec, and XMM  has a circular field of view with radius 15 arcmin 
and a spatial resolution of a few arcsec. INTEGRAL has an even wider circular field of view -- the
radius is 10 degree -- with a poorer spatial resolution (12 arcmin). There is therefore
no doubt that, for a typical LISA source at $z = 1$, GW observations
will provide information accurate enough to cover the interesting portion of the
sky with a single observation by one of (or all) the former instruments.
At $z = 1$, the typical size of a galaxy is $\simeq 3$ arcsec and the typical
size of a galaxy cluster is $\simeq 2$ arcmin. In a very few lucky occasions LISA will be able
to detect a source with $\Delta\Omega_N \approx 3\times 10^{-3}\,{\rm deg}^2$, but more typically 
the angular resolution is $\Delta\Omega_N \approx 0.3\,{\rm deg}^2$, and could be much worse for
a few detections ($\Delta\Omega_N \approx 5\,{\rm deg}^2$). It is therefore clear that for
the "high spatial resolution" GW detections, only $\approx 3$ galaxy clusters will fall
in the LISA error box, but more typically the number of clusters will be $\approx 300$. If there is 
still a small (but not negligible) chance of identifying the galaxy cluster hosting
a GW source, the chances of pin-pointing the host galaxy seem to be very bleak (the former numbers
need to be multiplied by a factor $\sim 10^3$ for galaxies).

\section{Conclusion}
\label{sec:concl}

We have considered LISA observations of massive black hole binary systems in the
final stage of in-spiral for rapidly rotating black holes. We have restricted our analysis 
to comparable mass objects in circular orbit, modelling the radiation at the restricted
post$^{1.5}$-Newtonian order assuming simple precession.
We have derived read-to-use analytical expressions for the signal registered at the
detector output, both in the time and frequency domain, and have determined the 
mean-square-errors associated to the parameter measurements for equal mass $10^6\Ms$ binary
systems at $z = 1$ for selected spin parameters and a wide range of source
locations and orientations. Our analysis clearly shows that 
the presence of spins reduces (by orders of magnitude, for some parameters) 
the errors with which the source parameters are measured. LISA is therefore a more
powerful telescope than previously thought if spins play a significant role.
The main shortcoming of our analysis is the limited region of the parameter space that
we have been able to explore, which is entirely due to our limited computational resources.
Such analysis is currently in progress~\cite{Vecchioetal03}. Of particular interest
is the exploration of the parameter space in the case where the masses are
not comparable, {\em i.e.} the mass ratio is $\sim 0.1$ or 
smaller. Our present understanding of the relevant astrophysical scenarios 
suggests, in fact, that the formation of unequal mass binaries should be regarded
as the rule rather than the exception. When the mass ratio increases,
the number of precession cycles detected at the LISA output will increase as well, cf.
Table~\ref{tab:Nprec} and Eq.~(\ref{Np1}). It is therefore conceivable that 
precession will be even more
effective in breaking the degeneracy among the parameters, in particular the two 
masses. However the amount of intrinsic rotation $S$ shall also play a crucial 
role, as the number of precession cycles is proportional to $S$.
Preliminary results suggest that for $m_2/m_1 = 0.1$ and rapidly rotating systems 
($S/M^2 \approx 0.9$) the estimation of the mass parameters and the
angular resolution improve by a factor of a few, with respect to the equal mass
case, provided that the signal-to-noise 
ratio is not strongly affected. In general, in fact, unequal mass systems will be
detected at a lower SNR, so that precession compensates for the loss of SNR. 
The latter 
eventually dominates the effect of rotation  when $m_2/m_1$ decreases further 
and compromises the accuracy with which parameters can be measured.

\begin{acknowledgments}

We would like to thank C. Cutler and A. Sintes for several discussions about LISA
parameter estimation. We would also like to thank L. Jones and I. Stevens for
discussions about high-redshift surveys of galaxies and clusters. 

\end{acknowledgments}


\begin{thebibliography}{}

\bibitem{lisa_ppa}
P.~L.~Bender et al., 
{\it LISA Pre-Phase A Report; Second Edition}, MPQ 233 (1998). P.~L.~Bender
et al., {\it LISA -- System and Technology Study Report},
ESA-SCI(2000)11, (2000). http://www.lisa.jpl.gov

\bibitem{CT02}
C.~Cutler and K.~S.~Thorne, {\em An Overview of Gravitational-Wave Sources},
to appear in Proceedings of GR16 (Durban, South Africa, 2001), gr-qc/0204090.

\bibitem{Magorrianetal98}
J.~Magorrian et al, AJ {\bf 115}, 2285 (1998).

\bibitem{Richstoneetal98}
D.~Richstone et al., Nature {\bf 395}, A14 (1998).

\bibitem{Genzeletal97}
R.~Genzel, A.~Eckart, T.~Ott, and F.~Eisenhauer,  
Mont. Not. Roy. Astron. Soc. {\bf 291}, 219 (1997).

\bibitem{Miyoshietal95}
M.~Miyoshi, J.~Moran, J.~Herrnstein, L.~Greenhill, N.~Nakai,
P.~Diamond, M.~and Inoue, Nature {\bf 373}, 127 (1995).

\bibitem{Maoz98}
E.~Maoz, Astrophys. J. {\bf 494}, L181 (1998).

\bibitem{Fanetal01}
X.~Fan et al, AJ {\bf 122}, 2833 (2001).
 
\bibitem{WL02}
J.~S.~B.~Wyithe and A.~Loeb, astro-ph/0206154.

\bibitem{MF01}
D.~Merritt and L.~Ferrarese, ApJ {\bf 547}, 140 (2001).

\bibitem{BBR80}
M.~C. Begelman, R.~D.~Blandford, and M.~J.~Rees, Nature {\bf 287}, 307 (1980).

\bibitem{VCP94}
A.~Vecchio, M.~Colpi and A.~Polnarev, ApJ {\bf 433}, 733 (1994)

\bibitem{RR95}
M.~Rajagopar and R.~W.~Romani, ApJ {\bf 446}, 543 (1995).

\bibitem{MM01}
M.~Milosavljevic and D.~Merritt, ApJ {\bf 563}, 34 (2001).

\bibitem{Yu02}
Q.~Yu, MNRAS {\bf 331}, 935 (2002).

\bibitem{GR00}
A.~Gould and H.~Rix, ApJ {\bf 532}, L29 (2000)

\bibitem{AN02}
P.~J.~Armitage and P.~Natarajan, ApJ {\bf 567}, L9 (2002)

\bibitem{Haehnelt94}
M.~G.~Haehnelt, Mont. Not. Roy. Astron. Soc. {\bf 269}, 199 (1994).

\bibitem{Haehnelt98}
M.~G.~Haehnelt, in {\it Laser Interferometer Space Antenna}, 
ed. W. M. Falkner (AIP Conference Proceedings 456), pp. 45-49 (1998).

\bibitem{Vecchio97}
A.~Vecchio, Class. Quant. Grav. {\bf 14}, 1431 (1997).

\bibitem{MHN01}
K.~Menou, Z.~Haiman and V.~K.~Narayanan. ApJ {\bf 558}, 535 (2001)

\bibitem{Menou02}
K.~Menou 2002, astro-ph/0301397

\bibitem{JB02}
A.~H.~Jaffe and D.~C.~Backer, astro-ph/0210148 (2002).

\bibitem{WL03}
J.~S.~B.~Wyithe and A.~Loeb, astro-ph/0211556.

\bibitem{Pursimoetal00}
T.~Pursimo et al, Astron. Astrophys. Suppl {\bf 146}, 141 (2000).

\bibitem{Cutler98}
C.~Cutler, Phys. Rev. D {\bf 57}, 7089 (1998).

\bibitem{SV00}
A.~M.~Sintes and A.Vecchio, in
{\it Gravitational Waves -- Third Amaldi Conference},
Ed. S.~Meshkov (American Institute of Physics), pp. 403-404 (2000).

\bibitem{Sintes03}
A.~M.~Sintes at al, in preparation

\bibitem{MH02}
T.~A Moore and R.~W.~Hellings, R.~W., Phys. Rev D. {\bf 65}, 062001 (2001).

\bibitem{Hughes02}
S.~Hughes, MNRAS {\bf 331}, 805 (2002)

\bibitem{Seto02}
N.~Seto, PRD in press.

\bibitem{HB03}
S.~A.~Hughes and R.~D.~Blandford, Ap.J. Lett (in press), astro-ph/0208484

\bibitem{ACST94}
T. A. Apostolatos, C. Cutler, G. S. Sussman and K. S. Thorne
Phys. Rev. D {\bf 49}, 6274 (1994).

\bibitem{Kidder95}
L.E. Kidder, Phys. Rev. D {\bf 52}, 821 (1995).

\bibitem{VC98}
A.~Vecchio and C.~Cutler, in 
{\it Laser Interferometer Space Antenna}, ed. W. M. Falkner (AIP Conference
Proceedings 456), pp. 101-109 (1998).

\bibitem{Vecchio00}
A.~Vecchio, in 
{\it Gravitational Waves -- Third Amaldi Conference}, 
Ed. S.~Meshkov (American Institute of Physics), pp. 238-247 (2000).


\bibitem{Vecchioetal03}
A.~Vecchio et al, in preparation.

\bibitem{frame}
We would like to caution the reader of the different notation used in this
paper with respect to~\cite{Cutler98}:
the reference frames $(x,y,z)$ and $(x',y',z')$ of this paper coincide,
respectively, with the frames $(\bar{x},\bar{y},\bar{z})$ and $(x,y,z)$ of~\cite{Cutler98}.


\bibitem{Blanchet02}
L.~Blanchet, Living Review  in Relativity 2002-3
http://www.livingreviews.org/Articles/Volume5/2002-3blanchet/index.html
(2002)

\bibitem{FH98}
E.~E.~Flanagan and S.~Hughes, Phys. Rev. D {\bf 57}, 4535, 1998.

\bibitem{Peters64}
P.~C.~Peters, Phys. Rev. {\bf 136}, B1224 (1964). 

\bibitem{Thorne87}
K.S. Thorne, in {\it 300 Years of Gravitation},
edited by S. W. Hawking and W. Israel (Cambridge University Press,
Cambridge, England, 1987), pp. 330-458.

\bibitem{BIWW96} L. Blanchet,  B.R. Iyer, C.M. Will and A.G.
Wiseman, Class. Quantum. Gr. {\bf 13}, 575, (1996).

\bibitem{Weinberg}
S. Weinberg  {\it Gravitation and Cosmology: Principles and
Applications of the General Theory of Relativity} (New York: Wiley)
(1972)

\bibitem{BDIWW95} L. Blanchet, T. Damour, B.R. Iyer, C.M. Will and A.G.
Wiseman, Phys. Rev. Lett. {\bf 74}, 3515 (1995).


\bibitem{BO75}
B. M. Barker and R. F. O'Connell, Phys. Rev. D {\bf 12}, 329 (1975).

\bibitem{TH}
K.S. Thorne and J. B. Hartle, Phys. Rev. D {\bf 31}, 1815 (1985).


\bibitem{WZ62}
L.~A.~Wainstein and V.~D.Zubakov, {\em Extraction of signals from noise},
Dover, New York (1962).

\bibitem{Helstrom68} 
C.W.~Helstrom, {\em Statistical Theory of Signal Detection},
2nd edition, Pergamon Press, London (1968).

\bibitem{NV98}
D. Nicholson and A.~Vecchio, Phys. Rev. D {\bf 57}, 4588 (1998).

\bibitem{Finn92}
L.~S.~Finn, Phys. Rev. D {\bf 46}, 5236 (1992).

\bibitem{CF94} 
C.~Cutler and E.E.~Flanagan, Phys. Rev. D {\bf 49}, 2658 (1994)

\bibitem{Hilsetal90} 
D.~Hils, P.~L. Bender, and R.~F. Webbink, Ap.~J. {\bf 360}, 75 (1990).

\bibitem{BH97}
P.~L.~Bender and D.~Hils, Class. and Quantum Grav. {\bf 14}, 1439 (1997).

\bibitem{norm}
In this paper we are mainly concerned with a comparison of the errors obtained
without considering spins with the ones that are given when spins are included
in the signal.
It is therefore clear that the actual redshift (distance) of the fiducial source
and the choice of the expected noise spectral density (instrumetal noise
and confusion noise) have little interest, as far as the same values are used
in the two cases.




\end{thebibliography}
\end{document}